\documentclass{nature}

\usepackage{bm}
\usepackage{amsmath}
\usepackage{amssymb}
\usepackage[textsize=tiny]{todonotes}
\usepackage{lineno}

\renewcommand{\eqref}[1]{(\ref{#1})}
\newcommand{\figref}[1]{Fig.~\ref{#1}}

\title{Classical nucleation and growth of DNA-programmed \\ colloidal crystallization}

\author{Alexander Hensley$^{1}$, William M. Jacobs$^{2,*}$ \& W. Benjamin Rogers$^{1,*}$}

\begin{document}

\maketitle

\vspace{-0.5 cm}

\begin{affiliations}
 \item Martin A. Fisher School of Physics, Brandeis University, Waltham, MA USA, 02453
 \item Department of Chemistry, Princeton University, Princeton, NJ USA, 08544
\end{affiliations}


\spacing{1}
\begin{abstract}

  DNA-coated colloids can self-assemble into an incredible diversity of crystal structures, but applications of this technology are limited by poor understanding and control over the dynamical crystallization pathways.
  To address this challenge, we use microfluidics to quantify the self-assembly dynamics of DNA-programmed colloidal crystals, from thermally-activated nucleation through reaction-limited and diffusion-limited phases of crystal growth.
  Our detailed measurements of the temperature and concentration dependence of the kinetics at all stages along the crystallization pathway provide a stringent test of classical theories of nucleation and growth.
  After accounting for the finite rolling rate of micrometer-sized DNA-coated colloids, we find that modified versions of these classical theories quantitatively predict the absolute nucleation and growth rates.
  We conclude by applying our model to design and demonstrate protocols for assembling large single crystals, including crystals with pronounced structural coloration, an essential step in the creation of next-generation functional materials from colloids.

\end{abstract}
\spacing{2}

\newpage

\section{Introduction}
By encoding specific short-range interactions, DNA molecules grafted to colloidal particles can be used to direct the self-assembly of complex, crystalline materials \cite{jones2015programmable, Rogers2016, laramy2019crystal}.
This general approach to crystal engineering is a triumph of synthetic self-assembly and has yielded a vast diversity of crystal structures with programmable stoichiometries, composition, and crystallographic symmetries from both nanometer- \cite{Mirkin2008Nature, Oleg2008Nature, jones2010dna, macfarlane2011nanoparticle, zhang2013general, auyeung2014dna, liu2016diamond} and micrometer-scale particles \cite{Crocker2012NComm,rogers2015programming,Wang2015,Wang2017NComm,ducrot2017colloidal,fang2020two,he2020colloidal}.
While the breadth of such structures has increased dramatically over time, a detailed understanding of the self-assembly pathways has remained elusive.
New experimental methodologies are thus needed to test theoretical models of the pathways governing crystallization and, ultimately, to achieve control over the dynamics of self-assembly.

Colloidal crystals are widely believed to self-assemble via classical nucleation and growth, following dynamical pathways analogous to those of atoms and simple molecules.
According to classical nucleation theory (CNT), a crystalline nucleus spontaneously forms from a metastable fluid by surmounting a free-energy barrier \cite{oxtoby1992homogeneous}.
Subsequent growth then occurs by the addition of free particles to the nucleus.
A central challenge in programmable self-assembly of colloids is to understand whether this framework quantitatively describes the crystallization dynamics of micrometer-sized colloidal particles. On the one hand, colloidal particles can be thought of as ``model atoms'' that interact via an effective interaction potential that is averaged over all of the molecular degrees of freedom\cite{Rogers2011PNAS}. On the other hand, the effective interaction arises from the transient formation and rupture of very real DNA duplexes that link neighboring particles together, whose kinetics may dramatically influence the rates of local rearrangements within a colloidal assembly \cite{Wang2015,Miranda2018SoftMatter}.  
Such dynamical considerations are crucially important, as numerous examples of colloidal self-assembly have shown that the thermodynamically stable phase that one would predict on the basis of the effective interactions alone is not always accessible, and that these systems are prone to becoming arrested as a colloidal gel instead \cite{Dreyfus2010,Michele2013}.

Here we quantify the nucleation and growth dynamics of DNA-programmed crystallization in a binary mixture of colloidal particles.
By monitoring the self-assembly of hundreds of isolated crystals simultaneously, we show that a modified version of classical nucleation theory---which takes into account the finite rate at which bound particles `roll' over one another at the crystal interface---quantitatively describes the observed temperature and concentration dependence of the nucleation barrier, as well as the absolute nucleation rate.
Furthermore, our model of the rolling-mediated attachment kinetics successfully captures the dynamics of the initial reaction-limited phase of crystal growth, which occurs before large crystals ultimately enter into a deterministic, diffusion-limited growth regime.
With this understanding of the crystallization dynamics, we accurately predict the extremely narrow temperature window---less than $0.1^\circ\mathrm{C}$---in which large, faceted single crystals can be grown.
We then use this knowledge to design and demonstrate a protocol for assembling millions of single crystals of DNA-coated particles that exhibit a pronounced photonic response, thereby overcoming a critical hurdle to using DNA-programmed assembly to build optical metamaterials \cite{he2020colloidal,ross2015nanoscale,sun2018design,Park2019ACS}.

\section{Results}

\begin{figure*}
    \centering
    \includegraphics[width=\linewidth]{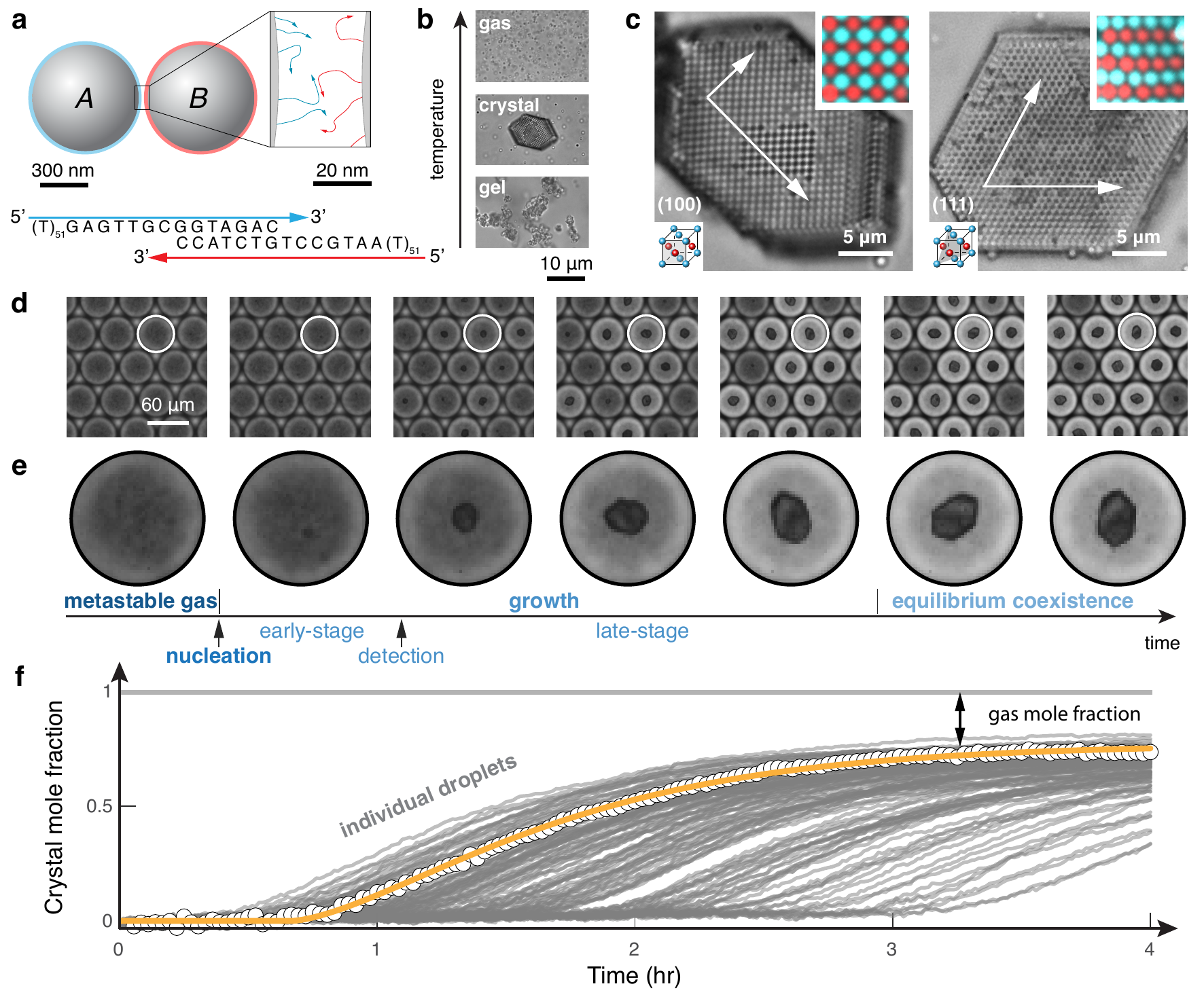} \spacing{1}
    \caption{ \textbf{DNA-coated colloids follow a dynamic pathway to crystallization characterized by three distinct regimes: nucleation, growth, and equilibrium coexistence.} (\textbf{a}) A binary mixture of 600-nm-diameter colloidal particles that interact by direct hybridization of complementary DNA sequences. (\textbf{b}) The resultant interactions are temperature dependent, forming a colloidal gas phase at high temperature, an ordered crystal phase at intermediate temperature, and a disordered gel phase when quenched to low temperatures. (\textbf{c}) The binary mixture forms crystals with a copper-gold lattice structure. Brightfield and fluorescence micrographs show the (100) and (111) planes, as well as the crystal facets. (\textbf{d}) Micrographs of water droplets containing DNA-coated particles show the dynamics of nucleation and growth, as seen in the time series. (\textbf{e}) Time-lapse micrographs of a single droplet show the full dynamic pathway to crystallization, which proceeds through multiple stages: (i) nucleation from a metastable gas; (ii) early-stage growth; (iii) detection of a crystal followed by late-stage growth; and (iv) eventually equilibrium coexistence between crystal and gas. (\textbf{f}) We extract the crystal mole fraction as a function of time for each droplet individually. Points show data for a single droplet and the orange curve shows a model of growth described in the main text. Grey lines show trajectories of other droplets in the same experiment. Data are for a colloid concentration of 1\% (v/v).}
    \label{fig:intro}
\end{figure*}

We follow the full dynamic pathways to crystallization using optical microscopy and droplet-based microfluidics. Hybridization of complementary DNA strands grafted to colloidal particles (\figref{fig:intro}a) induces a short-range attraction whose strength can be tuned by adjusting the temperature (\figref{fig:intro}b). Just below the melting temperature, the colloids assemble into a binary alloy that is isostructural to copper-gold (CuAu-FCC, \figref{fig:intro}c) \cite{Crocker2012NComm}. By combining equal amounts of both particle species inside monodisperse, $100$-pL-volume droplets, we image and quantify hundreds of crystallization experiments running in parallel---one experiment per droplet  (\figref{fig:intro}d) \cite{Pound1952JACS,galkin1999direct,akella2014emulsion}. Because the number of free particles decreases as a crystal grows, the droplets become brighter after the initial nucleation event (\figref{fig:intro}e), enabling us to follow the entire dynamic evolution by quantifying the concentration of free particles, and thus the mole fraction of the crystal phase, from measurements of the transmitted intensity (\figref{fig:intro}f; see SI Section 5 for details). Importantly, we are able to track many instances of nucleation and growth within a single experiment, enabling precise quantification of the behavior of both individual crystals as well as the ensemble of many crystals \cite{Sear2014CrystEngCom}. Furthermore, this experiment can be repeated at many temperatures by heating the system to the gas phase and then quenching to a new temperature (see SI Sections 1--3 for experimental details).

The transition from a metastable, disordered gas to an ordered crystal is a complex dynamic process that follows a sequence of stages (\figref{fig:intro}e). First, we observe a metastable fluid at short times during which there are no visible stable nuclei. After some waiting time, which varies widely from droplet to droplet (\figref{fig:intro}f), we observe the spontaneous emergence of small crystallites. Next, the nucleated crystals grow in size as particles from the the gas phase adsorb to the growing crystal surface. Eventually, the crystals stop growing. The observation that crystals nucleate at a variety of times (\figref{fig:intro}f) suggests some underlying stochasticity and hints at the presence of a free-energy barrier between the gas and crystal phases. In contrast, following nucleation, crystal growth is consistent from droplet to droplet, suggesting that growth is nearly deterministic. Similarly, all crystals stop growing at the same crystal mole fraction, indicating that the crystals eventually equilibrate with a dilute gas phase. Thus, from a single experiment, we quantify the kinetics of both nucleation and growth, as well as the thermodynamic driving force, which we can then dissect to construct a quantitative model of the dynamic crystallization process. 

\begin{figure}
    \centering
    \includegraphics[width=\linewidth]{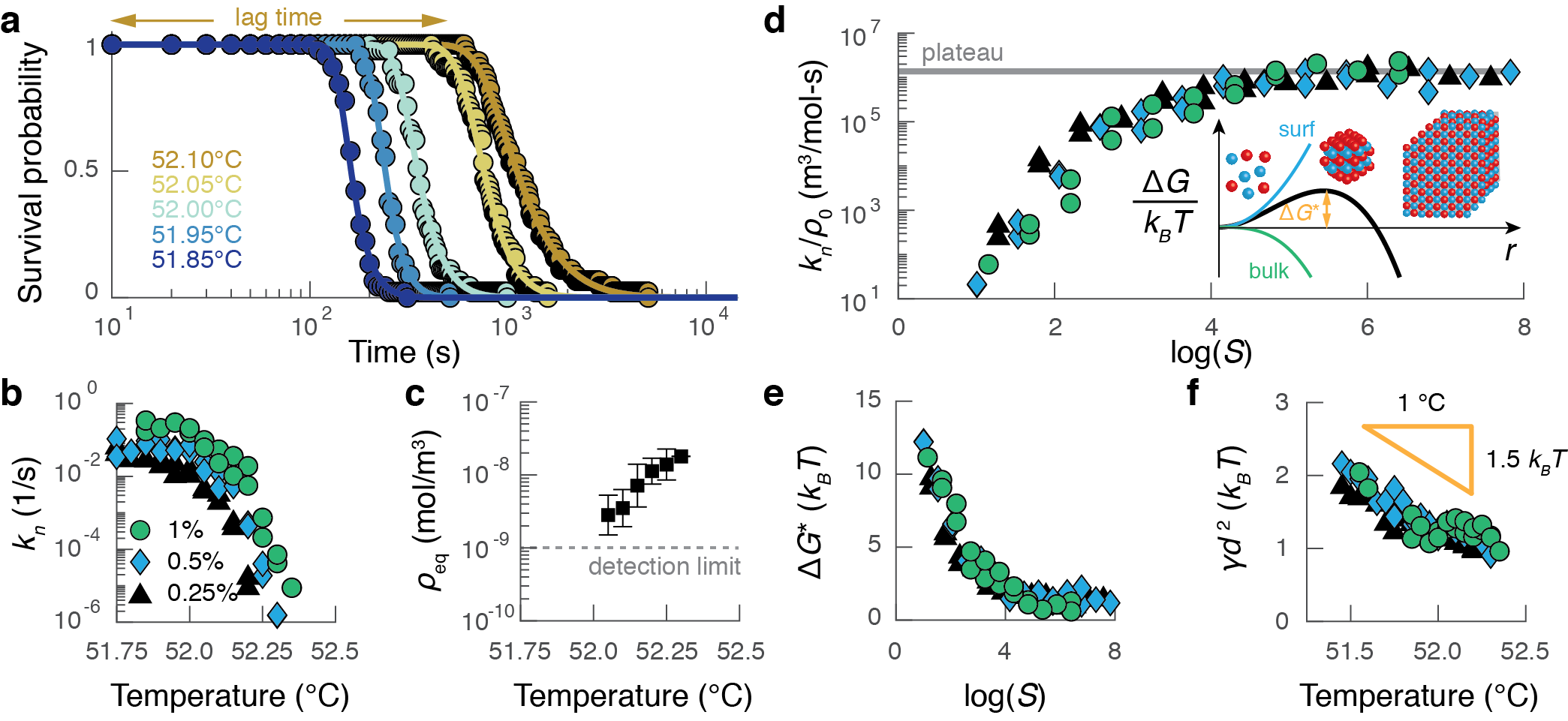} \spacing{1}
    \caption{\textbf{The kinetics of nucleation can be predicted by a modified version of classical nucleation theory.}   (\textbf{a}) The survival probability for a colloid concentration of 0.25\% (v/v) as a function of time for different temperatures (colors). Points show data and curves show fits of the model described in the main text. (\textbf{b}) The measured nucleation rate $k_{\text{n}}$ as a function of temperature at three different particle volume fractions: 0.25$\%$ (black triangles), 0.50$\%$ (blue diamonds), and 1.00$\%$ (green circles). At the lowest temperatures the nucleation rate plateaus; at higher temperatures the nucleation rate decreases super-exponentially with temperature. (\textbf{c}) Average measurements of the equilibrium gas density as a function of temperature. The error bars represent the standard deviation of the measurements for all three colloid concentrations and two independent experimental trials. (\textbf{d}) The nucleation rate divided by the initial colloid concentration as a function of the supersaturation $S$. The inset illustrates the free-energy barrier that results from the classical theory of nucleation. (\textbf{e}) The inferred barrier height $\Delta G^*$ as a function of supersaturation.  (\textbf{f}) The inferred surface tension $\gamma$ as a function of temperature for the experiments in \textbf{b}. The surface tension decreases with increasing temperature as expected given the temperature dependence of DNA hybridization; $d$ is the colloid diameter. }
    \label{fig:nucleation}
\end{figure}

To study the nucleation behavior, we measure the survival probability---the fraction of droplets that have not yet formed a crystal---as a function of time for a variety of temperatures (\figref{fig:nucleation}a) and three nominal colloid concentrations.
After accounting for random concentration variations between the droplets, which vary by roughly $\pm 5\%$ of the mean concentration, we find that the survival probabilities are characterized by an exponential decay at long times, suggesting that a single rate, $k_{\mathrm{n}}(\rho_0,T)$, controls nucleation at each temperature $T$ and concentration $\rho_0$.
This rate most likely describes homogeneous nucleation, since the colloidal particles are repelled from the oil--water interface by a polymeric surfactant and are nearly density-matched to the solvent, implying that sedimentation plays an insignificant role during the initial nucleation process.
We also observe a lag period with a soft shoulder at short times, which we attribute to the fact that the crystals must grow to a threshold size before they are detected by our image-analysis methods.
Indeed, the survival probability for each quench can be well fitted by a simple first-passage model to determine the nucleation rate $k_{\mathrm{n}}$ (\figref{fig:nucleation}b) and the mean lag time, $\tau_{\mathrm{lag}}$, which is discussed further below (see SI Section 7 for details).
Both of these quantities vary by several orders of magnitude within a temperature range of roughly $0.5^\circ\mathrm{C}$ for all three concentrations.

We analyze our measurements of the homogeneous nucleation rate within the framework of classical nucleation theory, which has been used to describe nucleation in a variety of systems, including molten metals, simple liquids, protein solutions, and colloidal suspensions~\cite{Palberg1999CondensedMatter}.
CNT predicts that a single free-energy barrier separates a metastable fluid from a globally stable crystal, resulting in a nucleation rate of the form $k_{\mathrm{n}} = k_{\mathrm{n},0}(\rho_0,T)\exp\left(-\Delta G^*/k_{\mathrm{B}}T\right)$, where $k_{\mathrm{n},0}(\rho_0,T)$ is the nucleation rate prefactor, $\rho_0$ is the initial gas number density, and $T$ is the absolute temperature.
The height of the nucleation barrier, $\Delta G^*$, is determined by a competition between the temperature-dependent interfacial free energy and the thermodynamic driving force for assembling the bulk crystal phase, $\Delta\mu$.
To determine $\Delta\mu$, we use the measured equilibrium gas number density, $\rho_\mathrm{eq}$, from each quench (\figref{fig:nucleation}c), to calculate the supersaturation, $S(T) \equiv \exp(-\Delta\mu/k_{\mathrm{B}}T) \simeq \rho_0/\rho_\mathrm{eq}(T)$.

Analyzing the temperature and concentration dependence of our measured nucleation rates yields estimates of the free energy barrier height and the surface free energy for DNA-programmed crystallization.
Plotting the measured nucleation rates as a function of the supersaturation reveals two distinct regimes predicted by CNT: a barrier-dominated regime at low supersaturation, and a temperature-independent plateau at high supersaturation, where the free-energy barrier is negligibly small relative to $k_{\mathrm{B}} T$ (\figref{fig:nucleation}d).
Assuming that the nucleus is roughly spherical and has the same crystallographic symmetry as the bulk crystal, CNT predicts a barrier height of the form $\Delta G^* = 16\pi\gamma(T)^3 / 3\rho_{\mathrm{c}}^2(\log S)^2$, where $\gamma(T)$ is the interfacial free energy density and $\rho_{\mathrm{c}}$ is the number density of the crystal.
Taking the plateau value of $k_n/\rho_0$ to be equal to the nucleation rate prefactor, we find that $\Delta G^*$ decreases from approximately 10~$k_{\mathrm{B}}T$ at the lowest supersaturations to near 0~$k_{\mathrm{B}}T$ above $\log{S}\approx 5$ (\figref{fig:nucleation}f).
These calculations suggest that the critical nucleus contains on the order of ten colloidal particles under the conditions in which the free-energy barrier is rate-limiting.
We highlight that this estimate of a 10-particle critical nucleus is consistent with homogeneous nucleation due to short-range attractions and strong driving forces ($\Delta\mu > k_{\text{B}} T$).
Furthermore, the surface free energies that we obtain from all temperatures and concentrations collapse onto a single curve, which decreases linearly with increasing temperature (\figref{fig:nucleation}e).
Importantly, both the magnitude and the temperature dependence of $\gamma(T)$ are consistent with independent estimates of the surface tension based on either the binding free energy between DNA-coated colloids~\cite{Rogers2011PNAS,lowensohn2019linker} or the equilibrium gas density shown in \figref{fig:nucleation}b (see SI Section 6 for details).
This agreement between experiment and calculation provides a strong justification for modeling nucleation with classical nucleation theory.

Our measurements of the nucleation rates at high supersaturation, where the nucleation rate is determined solely by the nucleation rate prefactor, reveal two additional interesting results.
First, the nucleation rate prefactor is a very weak function of temperature.
We hypothesize that any temperature dependencies are undetectable given the narrow temperature window of our experiment, which is only $0.5^\circ\mathrm{C}$ wide; for example, we estimate that the self-diffusion coefficient of the particles increases by only 1\% from 51.75--52.25$^\circ$C, which is below the precision of our measurements of the nucleation rate.
Second, and more surprisingly, the nucleation rate prefactor scales linearly with the initial gas density.
This observation contrasts with other examples of nucleation in which the nucleation rate prefactor is diffusion-limited and scales as $V_{\mathrm{drop}} \rho_0 D / \lambda^2 \propto \rho_0^{5/3}$, where $V_{\mathrm{drop}}$ is the droplet volume, $D$ is the self-diffusion constant, and $\lambda$ is the mean free path between particles in the gas phase \cite{Palberg1999CondensedMatter}.

The linear dependence of the nucleation rate prefactor on the initial gas density suggests that the pathway to forming a critical nucleus is fundamentally different for DNA-coated colloids, as compared to atoms, molecules, or other colloidal suspensions. We understand this unique functional dependence by considering the specific attachment kinetics for micrometer-sized DNA-coated particles. 
When a particle from the gas phase attaches to a crystalline nucleus, it must first roll on the surface of the cluster before settling at a metastable position within the emerging lattice.
This process can be slowed dramatically by the transient formation and rupture of DNA linkages \cite{Wang2015,Miranda2018SoftMatter}.
Given the characteristic time, $\kappa^{-1}$, for a colloid to roll a distance equal to its own radius, the rate of attachment of a single particle to the emerging crystalline lattice can be modeled as
\begin{equation}
  \label{eqn:katt}
  k_{\mathrm{att}}(S) = \frac{\kappa}{1+K(S)^{-1} + \kappa \lambda^2 / D},
\end{equation}
where $K(S)$ is the equilibrium constant for an adsorbed particle in a non-crystalline-lattice position.
When adsorbed single particles are highly unstable, the prefactor reduces to ${k_{\mathrm{n},0} \simeq \kappa K(1) V_{\mathrm{drop}} \rho_0}$, reproducing the linear dependence of the nucleation rate on the colloid density observed in our experiments. 
This model also predicts the absolute nucleation rate at high supersaturation to within an order of magnitude of our measurements using independent estimates of $\kappa \simeq 0.1~\mathrm{s}^{-1}$ \cite{Wang2015}, $D \simeq 10^{-12}~\mathrm{m}^2\mathrm{s}^{-1}$, and $K(1) \simeq 10^{-3}$ for a critical nucleus, providing further support of this interpretation (see SI Section 6 for details).
In contrast, the assumption of a diffusion-limited prefactor overestimates the nucleation rates by at least four orders of magnitude.
Therefore, it appears that while the interactions between particles can be accurately described by an effective potential that averages over the molecular degrees of freedom, capturing the dynamics of nucleation requires incorporating the effective friction that results from transient bridge formation, an effect that is exclusive to the colloidal length scale.

\begin{figure}
    \centering
    \includegraphics[width=0.5\linewidth]{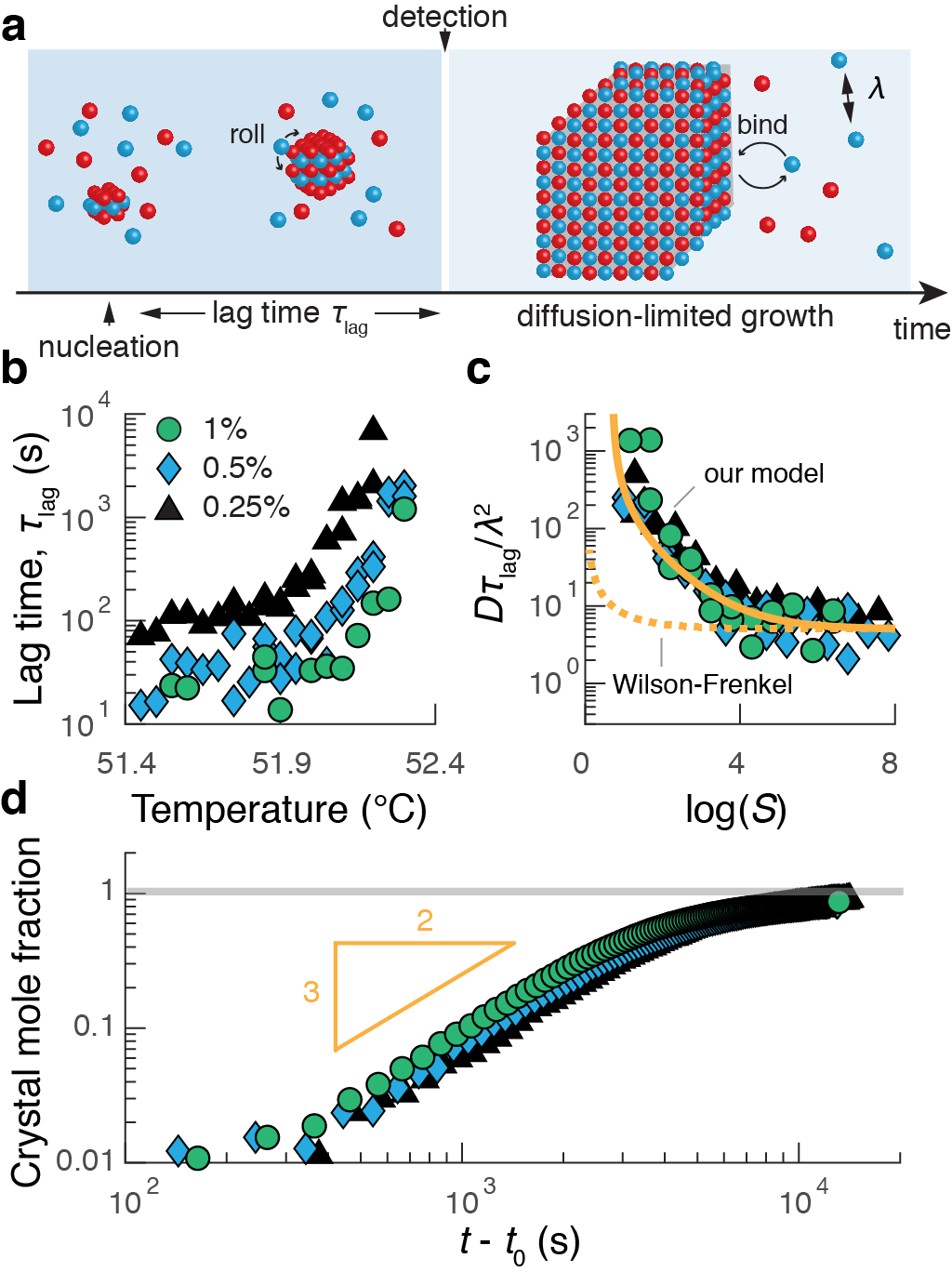} \spacing{1}
    \caption{\textbf{The growth kinetics are reaction-limited just after nucleation and become diffusion-limited as the crystal grows larger.} (\textbf{a}) An illustration of the stages of growth following the formation of a critcal cluster. (\textbf{b}) The inferred average lag time for early-stage growth $\tau_\textrm{lag}$ as a function of temperature for three colloid concentrations: 0.25\% (black triangles), 0.5\% (blue diamonds), and 1\% (v/v) (green circles). (\textbf{c}) The same lag times in (\textbf{b}) rescaled by the diffusion coefficient $D$ and the mean-free path $\lambda$ as a function of the measured supersaturation $S$. The solid orange curve shows a fit of SI Equation 22, which is described in the text; the dashed orange curve shows the predictions of the Wilson--Frenkel model\cite{saito1996statistical}. (\textbf{d}) The average crystal mole fraction as a function of the time since nucleation exhibits power-law growth with an exponent of roughly 3/2, consistent with predictions of diffusion-limited growth (see SI Section 6.4). Data in \textbf{d} are for temperatures at which mostly single crystals nucleate and grow: 52.15, 52.25, and 52.3$^\circ$C for 0.25\%, 0.5\%, and 1\% respectively. $t_0$ is the nucleation time for a given droplet.}
    \label{fig:growth}
\end{figure}

We now turn to analyzing the growth stage of the crystallization pathway.
We study the earliest stage of growth by analyzing the mean lag time between the formation of a critical nucleus and the moment that a post-critical cluster is detected (\figref{fig:intro}e, \figref{fig:growth}a).
Based on the resolution of our imaging setup and the specifics of our crystal-detection algorithm, we estimate that this smallest detectable cluster contains on the order of 50--200 colloidal particles.
From the survival probabilities shown in \figref{fig:nucleation}a, we find that the lag times vary over several orders of magnitude and are temperature- and concentration-dependent (\figref{fig:growth}b).
When rescaled by the characteristic timescale for diffusion-limited collisions and plotted against the thermodynamic driving force, the lag times collapse onto a single curve (\figref{fig:growth}c) that provides an independent test of our rolling-limited attachment model presented above in Equation~\eqref{eqn:katt}.
Specifically, the mean early-stage growth rate, which is proportional to $\tau_{\mathrm{lag}}^{-1}$, is predicted to have the approximate form ${D (1 - 2S^{-1}) / \lambda^2 [2 + S^{-1}K(1)^{-1}]}$ (see SI Section 6.4 for details).
Fitting this expression to the data in \figref{fig:growth}c, we obtain $K(1) \approx 0.01$, which supports our hypothesis that individual adsorbed particles are unstable at non-crystalline lattice positions and that early-stage growth is reaction-limited.
Moreover, this model also accurately accounts for the observed variation in lag times up to $\log(S) \approx 5$, whereas the standard Wilson--Frenkel law for crystal growth \cite{saito1996statistical} predicts a measurable supersaturation dependence only when $\log(S) \lesssim 1$ (\figref{fig:growth}c).

Once a crystal grows large enough, the situation changes and the growth dynamics become limited by the diffusion of particles to the crystal--vapor interface.
Assuming a roughly spherical crystal with radius $R$, our model predicts that growth enters this regime when $d^2/\lambda R \lesssim k_{\mathrm{att}}\lambda^2/D$, after which the crystal mole fraction should increase as the $3/2$ power of the time. Replotting our measurements of the crystal mole fraction versus the time since nucleation for three shallow quenches reveals a power-law dependence with an exponent of roughly 3/2, as predicted (\figref{fig:growth}d; see SI Section 6.4 for details). 
The growth rate then decreases exponentially as the crystal approaches its equilibrium size.
Fitting the late-stage growth data to the deterministic, diffusion-limited growth law (\figref{fig:intro}f)
\begin{equation}
  \frac{dN_{\mathrm{c}}(t)}{dt} \simeq 4\pi R(t) D_{\mathrm{eff}} \left[\rho_0 - \frac{N_{\mathrm{c}}(t)}{V_{\mathrm{drop}}} - \rho_{\mathrm{eq}}\right],
\end{equation}
where $N_{\mathrm{c}}$ is the number of particles in the crystal phase, we obtain an effective diffusion constant, $D_{\mathrm{eff}}$, that agrees quantitatively with predictions from the Stokes-Einstein equation \cite{theBigMan1905} in droplets with single crystals (see Fig.~SM15).
Taken together, these results demonstrate that our theoretical framework captures the functional dependences and the absolute rates of the distinct rate-limiting steps at all stages of growth in a self-consistent manner.

\begin{figure}
    \centering
    \includegraphics[width=0.5\linewidth]{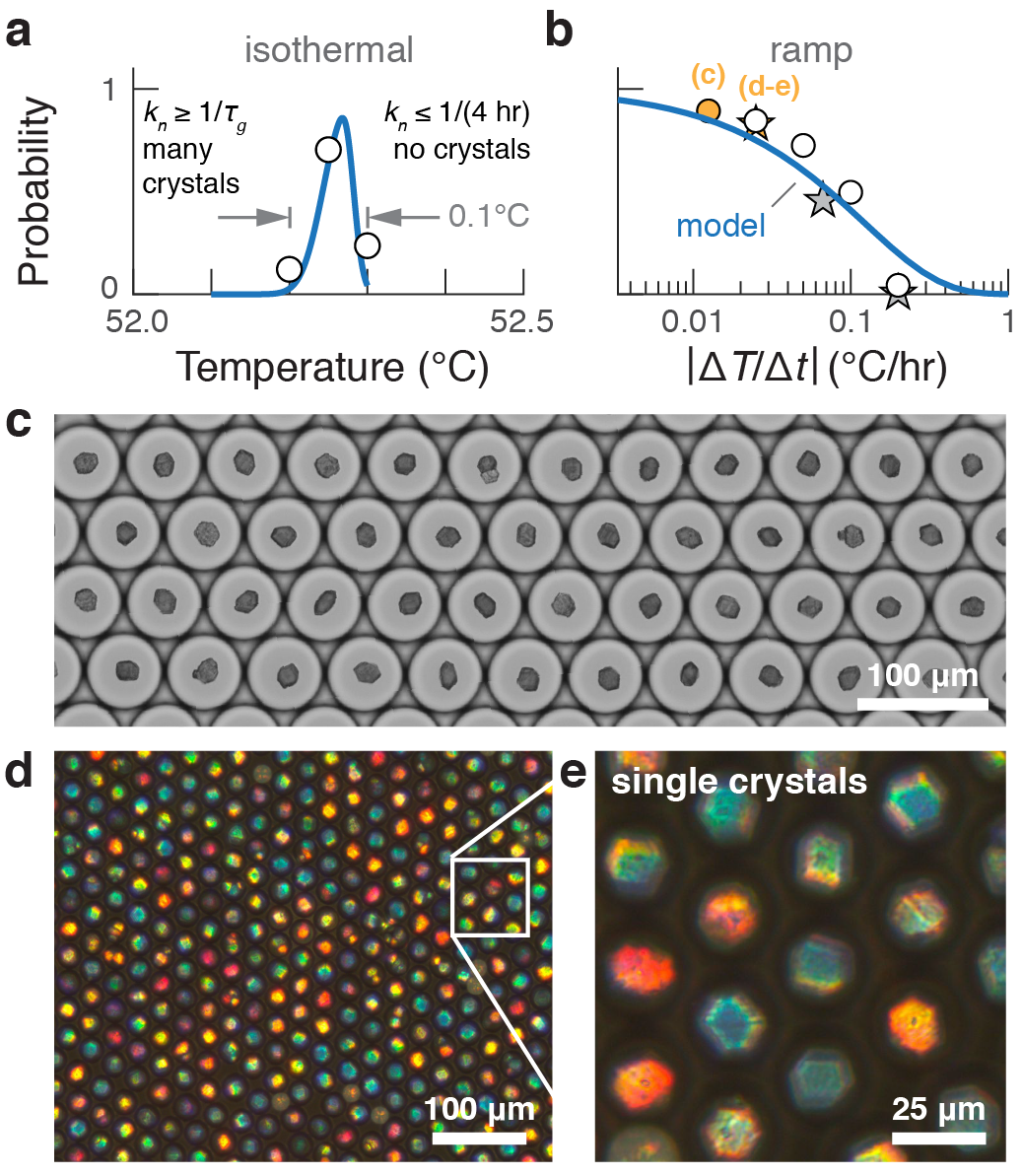} \spacing{1}
    \caption{\textbf{A temperature-ramp protocol yields large single crystals of DNA-coated colloids without the need for high-accuracy temperature control}. (\textbf{a}) The probability of forming a single crystal within a droplet held at a constant temperature for four hours as a function of temperature. Points show experimental measurements and the curve shows our model prediction, described in the text. (\textbf{b}) The probability of forming a single crystal within a drop that is cooled at a constant rate $\lvert \Delta T/\Delta t \rvert$. The circles show experimental measurements and the curve shows our model prediction for 600-nm-diameter particles at a concentration of 0.5\% (v/v); stars show experimental measurements for 400-nm-diameter particles at 1\% (v/v) (\textbf{c}) An optical micrograph of the temperature-ramp experiment for 600-nm-diameter particles at a ramp rate of 0.1$^\circ$C/8~hr (orange circle in \textbf{b}), showing that greater than 85\% of the droplets contain a single, well-faceted crystal. (\textbf{d}--\textbf{e}) Transmission optical micrographs showing that single crystals self-assembled from 400-nm-diameter colloids exhibit strong structural coloration when imaged through crossed polarizers. Crystals in \textbf{d}--\textbf{e} are formed from 1\% colloids at a ramp rate of 0.1$^\circ$C/4~hr (orange star in \textbf{b}). \label{fig:protocol}}
\end{figure}

We are now in a position to apply our quantitative understanding of the crystallization dynamics in order to grow large colloidal crystals.
Our specific aim is to assemble a single crystal per droplet with high probability, as this is an essential step in developing practical technologies based on colloidal crystallization.
In general, one should expect single crystals to form under conditions where nucleation is much slower than growth, since the addition of particles to a growing crystal lowers the supersaturation elsewhere in the droplet and thus reduces the rate of subsequent nucleation events\cite{dombrowski2010modeling, heymann2014room}.
Single crystals should therefore assemble when the nucleation rate, $k_{\mathrm{n}}$, is fast enough that nucleation occurs during the experiment, but slow enough that it is rare to observe two nuclei form within a time $\tau_{\mathrm{g}}$, which represents the typical growth time required to suppress additional nucleation events.
Because the nucleation rate drops rapidly with concentration, we reach $\tau_{\mathrm{g}}$ soon after entering the diffusion-limited phase of growth (see SI Section 8.1 for details).

Theoretical predictions suggest that forming single crystals with high probability using an isothermal protocol is intractable. \figref{fig:protocol}a compares the theoretical predictions of the probability of forming a single crystal per droplet, using $k_{\mathrm{n}}(\rho_0,T)$ and $\tau_{\mathrm{g}}(\rho_0)$ calculated from our model of the crystallization pathway, to the fraction of single crystals obtained at different temperatures.
While the close agreement between the theory and experiments confirms our intuition that the crystal morphology depends on a balance of nucleation and growth, we find that the temperature window within which we can grow single crystals with high probability is less than $0.1^\circ\mathrm{C}$ wide.
Unfortunately, sustained temperature precision and accuracy on this scale is difficult to achieve with conventional hardware, and consequently we observe polycrystals in most of our isothermal experiments.
An analysis of the full distributions of the number of crystals per droplet is presented in SI Fig.~SM16.

An alternative strategy is to perform the self-assembly in a temperature ramp in which the temperature decreases linearly with time \cite{auyeung2014dna}.
To predict the efficacy of this scheme, we compute the probability of forming a single crystal as a function of the ramp rate, $|\Delta T/\Delta t|$, assuming the same competition between $k_{\mathrm{n}}$ and $\tau_{\mathrm{g}}$ as above (\figref{fig:protocol}b; see SI Section 8.2 for details).
Encouragingly, our theory suggests that a ramp rate of $0.025^\circ\mathrm{C/hr}$ or slower is sufficient to guarantee a single crystal fraction of 75\% in our droplet system.
This prediction is borne out by a set of linear annealing experiments conducted at a range of ramp rates, which yield fractions of single crystals that closely match our predictions (\figref{fig:protocol}c; see SI Section 4 for details and SI Fig.~SM17 for additional experiments and predictions capturing the influence of the droplet volume).
Both the higher yield and the greater flexibility in choosing the ramp rate represent dramatic improvements compared to an isothermal protocol . 

Building on our ability to predict the efficacy of such a non-trivial experimental protocol from our quantitative understanding of the crystallization dynamics, we conclude by demonstrating a ramp protocol that produces an array of photonic crystals from DNA-coated colloids, thus realizing a longstanding goal of programmable self-assembly.
Assuming that reducing the particle diameter by 33\% minimally affects the parameters in our model, we choose a ramp rate that is predicted to yield primarily single crystals from 400-nm-diameter particles at a concentration of 1\% (v/v). Figure~\ref{fig:protocol}d--e show representative micrographs of the crystals that form. As predicted, 82\% of the droplets contain single crystals, which are each assembled from about 30,000 particles. Most strikingly, the crystals exhibit pronounced structural coloration, a photonic property that arises from the precise, periodic arrangement of the wavelength-sized colloidal particles. To the best of our knowledge, this is the first experimental demonstration of DNA-programmed assembly of photonic single crystals at optical length scales, an accomplishment that is enabled only through our detailed understanding of the dynamic pathways that govern crystallization.

\section{Discussion}

Our findings from these experiments are broadly two-fold.
First, we have demonstrated that the complete crystallization pathway can be understood in terms of classical theories of nucleation and growth, provided that a model of rolling-limited attachment kinetics is included to account for the finite rates of formation and rupture of the DNA linkages.
With this modification, our model predicts both the absolute nucleation and growth rates to within an order of magnitude of their measured values, a level of agreement between theory and experiment that stands in stark contrast with previous attempts to describe colloidal systems using CNT \cite{Weitz2001Science}.
Therefore, because of the large dynamic range of our measurements, our ability to suppress heterogeneous nucleation by eliminating impurities, and our ability to account for the relevant kinetics across multiple length scales, we believe that our experiments are among the most direct tests of classical nucleation theory to date in any molecular or colloidal system.
Furthermore, our results establish the first precise measurements of the temperature and concentration dependencies of the nucleation barrier, surface tension, and growth laws for micrometer-sized DNA-grafted colloids.
In particular, our quantification of the strong temperature dependence of the nucleation barrier may explain why forming large faceted crystals with these particles has been historically challenging \cite{Rogers2016}.
We have also applied these insights to predict the morphologies of crystals that form under various conditions and experimental protocols, culminating in the assembly of millions of large single crystals that exhibit pronounced structural colloration.
These achievements point the way towards the rational design of experimental protocols for guiding DNA-programmed colloidal self-assembly.

Second, our results hint at further practical applications of forming colloidal crystals in droplets.
Unlike in a bulk system, where the large variation in nucleation times leads to a broad distribution of crystal sizes, droplets can be used to grow millions of well-faceted single crystals with a specified, uniform size.
Our ability to predict this behavior suggests that droplets may be used to selectively self-assemble different crystal morphologies and sizes, following a size-limiting mechanism similar in spirit to the finite-pool mechanism of self-limiting assembly within living cells \cite{phillips2012physical}.
Furthermore, by using non-invasive methods to permanently crosslink the DNA-coated colloids once crystallized \cite{Feng2013NatMat,Lee2019} and then dissolving the droplet interface, it may be possible to use droplet-nucleated crystals to seed the continued growth of larger single crystals in bulk \cite{Allahyarov2015NComm}.
Finally, we note that other methods to soften the sharp temperature dependence of the nucleation rate, such as adding free DNA oligomers to compete with the binding of the grafted DNA strands \cite{rogers2015programming}, could be employed to increase the maximum temperature ramp rate for growing single crystals, thereby shortening the duration of the assembly process.
Taken together, our results promise that the long-standing goals of programming the complete self-assembly pathway to prescribed crystal structures \cite{mirkin1996dna, laramy2019crystal}, and then extending this technology to build more complex structures \cite{jacobs2016self}, may finally be within reach.

\section*{Methods}
\noindent \textbf{Synthesizing DNA-coated colloids}
Colloidal particles are functionalized with DNA using a combination of strain-promoted click chemistry and physical grafting, following a modified version of the methods described by Pine and co-workers\cite{Pine2015ChemMater}. In brief, polystyrene-block-poly(ethylene oxide) (PS-b-PEO) copolymers are functionalized with an azide group, this azide-modified PS-b-PEO is adsorbed onto the surface of the polystyrene colloidal particles, and then DBCO-modified DNA is attached to the PS-b-PEO via click chemistry. In detail, 100~mg of PS-b-PEO $(M_w=6500$ and $3800$~g/mol, Polymer Source, Inc.) is placed into a washed vial. 2~ml of dichloromethane (DCM, Sigma-Aldrich) and 42~$\mu$L triethylamine (TEA, Sigma-Aldrich) are then added to the vial and stirred with a stir bar until dissolved. The solution is  stirred in an ice bath for 15 minutes and then 23.5~$\mu$l of methanesulfonyl chloride (471259, Sigma-Aldrich) is added. The vial is covered with parafilm and stirred in an ice bath for 2 hours, then at room temperature for 22 hours. The solution is  dried in a Falcon tube in a vacuum desiccator for 6 hours. A mixture of 10~ml methanol and 243~$\mu$l of 37~$\%$ hydrochloric acid is poured into the tube with dried PS-b-PEO, vortexed, and then placed in a freezer to precipitate. The tube is centrifuged at 2~$^\circ$C at 4,500~g and the pellet is dissolved with 3~ml methanol.  40~ml of diethyl ether is added and the tube is placed back in the freezer to precipitate and be washed again. This process is repeated and then the pellet is dried in a desiccator overnight. 10~mg of sodium azide is added to another washed vial and dissolved in 2~ml dimethyl formamide with a stir bar. The desiccated PS-b-PEO pellet is then added to this vial and stirred in a 65~$^\circ$C oil bath for 24 hours. The contents of the vial are poured into a Falcon tube and washed four times with methanol and diethyl ether, like before, and then desiccated overnight. The dried PS-PEO-N$_3$ pellet has a molar mass of 10,342~g/mol and is suspended to 1~mM in deionized (DI) water.

Polystyrene colloids are washed five times in DI water via centrifugation at 4500~g for 10 minutes and suspended to 10$\%$(v/v). 167~$\mu$l of 1~mM PS-PEO-N$_3$, 33~$\mu$l DI water, 160~$\mu$l tetrahydrofuran, and 40~$\mu$l of the washed colloids are added to an Eppendorf tube. For dyed particles, 167~$\mu$l 1~mM PS-PEO-N$_3$, 73~$\mu$l DI water, 140~$\mu$l tetrahydrofuran, 3~$\mu$l dyed toluene (dyed with either Pyromethene 546 or Pyromethene 605 at 50$\%$ saturation), and 40~$\mu$l washed colloids are added to an Eppendorf tube. The Eppendorf tube is placed on a shaker plate for 30 minutes. The solution is then split into 4 separate Eppendorf tubes and filled to roughly 600~$\mu$l with DI water and left for an hour. The particles are then washed 5 times in DI water as above, recombined, and then suspended to a particle volume fraction of 1$\%$(v/v). For each DNA species, 10~$\mu$l DBCO DNA at 100~$\mu$M in DI water, 40~$\mu$l particles, and 150~$\mu$l 1xTE/1M NaCl/0.05$\%$ (by weight) F127 are added to an Eppendorf tube and rotated in a 65~$^\circ$C oven for 24 hours. The particles are washed 5 times in DI water via centrifugation as above.

\noindent \textbf{Fabricating the microfluidic device} The microfluidic drop-maker is fabricated via standard photolithographic techniques. A glob of SU8 (SU-8 2075, MicroChem) roughly the size of a quarter is poured onto a silicon wafer (3-76-024-V-B, Silicon Materials Inc.). The wafer is then spun at 500~rpm with a spin coater at a ramp rate of 100~rpm/sec for 5 seconds, and then 1500~rpm at a ramp rate of 300~rpm/sec for 60 seconds, which leads to a thickness of around 80~$\mu$m. Next, the wafer is placed onto a 65~$^\circ$C hot plate for 3 minutes and then a 95~$^\circ$C hot plate for 5 minutes. A photomask (Output City) with the pattern of our microfluidic device is placed on top of the wafer, which is then moved to a Manual Mask Aligner System (ABM-USA) and  exposed to UV light for 46 seconds. The mask is removed and the wafer is washed with isopropanol and propylene glycol methyl ether to remove the undeveloped photoresist. The wafer is then dried with an air brush and placed on a 65~$^\circ$C hot plate for 3 minutes and a 95~$^\circ$C hot plate for 20 minutes. Next, the wafer is placed in a glass Petri dish with PGME and shaken back and forth for 10 minutes to remove any photoresist. Finally, the wafer is sprayed with isopropanol and dried with an air brush.

The master is a negative of the actual device and acts as a mold. 30~g of polydimethylsiloxane (PDMS) and 3~g crosslinker (1673921, Dow Chemical Company) is mixed using a Thinky AR-250 planetary centrifugal mixer for 3 minutes. A plastic Petri dish is lined with aluminum foil and the microfluidic-device master is placed face up in the dish. The mixed PDMS is then poured onto the master and placed in a vacuum desiccator for 30 minutes to remove any bubbles from the solution. The dish is placed in a 70~$^\circ$C oven overnight. The wafer is removed from the dish, the foil is peeled off, and a hobby knife is used to cut away the excess PDMS and separate it fully from the master. A coring tool (69039-07, Electron Microscopy Sciences) is then used to punch holes into all the device inlets and outlets. A glass slide (2947-75X50, Corning) and the PDMS chip are placed into an oxygen plasma cleaner (Zepto, Diener electronic) for 45 seconds. The PDMS chip is then laid down onto the glass slide and held with uniform pressure for 30 seconds.

\noindent \textbf{Droplet making} Syringe pumps are used to operate the microfluidic device to produce monodisperse droplets containing colloidal suspensions. The channels of the microfluidic device are made hydrophobic by flushing them with Aquapel (B004NFW5EC, Amazon), leaving it for 30 seconds, and then flushing them out again with air to remove the Aquapel. The channels are flushed with HFE-7500 oil (3M) and then air. Flow rates are controlled by three 3-ml syringes independently with syringe pumps (98-2662, Harvard Apparatus) connected to the device via tubing (06417-11, Cole-Palmer) that is slightly larger in the diameter than the holes to ensure a snug fit. HFE-7500 with 2\% RAN fluorosurfactant (008-FluoroSurfactant-5wtH-20G, RAN Biotechnologies) is fed into the oil inlet, 1~M NaCl in 1xTE buffer is fed into one aqueous inlet, and DNA-coated particles suspended at twice the desired volume fraction in DI water are fed into the other aqueous inlet. The particles are loaded into the tube by using a reverse flow rate and never enter the syringe body. Flow rates of 800~$\mu$l/hr for the oil phase and 400~$\mu$l/hr for each of the aqueous phases are used to obtain droplets with diameters of roughly 60~$\mu$m which are collected in an Eppendorf tube via outlet tubing.

\noindent \textbf{Sample chamber} Sample chambers are comprised of a rectangular capillary filled with colloidal suspension that are glued to a glass coverslip. A 100-$\mu$m tall, 2-mm-wide glass rectangular capillary (5012, VitroCom) is cut to 3~cm in length with a glass scoring pen and held suspended in place with a pair of clamping tweezers. Approximately 2--3~$\mu$l of the droplet emulsion is transferred into the capillary via a micropipette that has been snipped at the tip to have a wider inlet. HFE 7500 with 2$\%$ RAN is used to fill the rest of the volume. UV glue is applied in a thin 3~cm line on a glass slide that the capillary is then placed onto and gently pressed flat. UV glue is then used to seal the two sides of the capillary tube, taking care not to introduce bubbles inside the capillary. The capillary and slide are placed under a UV lamp for 30 seconds, and then foil is used to cover all but the glue at the ends of the capillary to avoid UV damage of the DNA-coated particles. The sample is placed capillary-side down on an acrylic frame with a rectangular hole with a small amount of immersion oil fixing the slide to the acrylic. A drop of immersion oil is placed on the glass slide and a  PID-controlled Peltier unit with a central hole and sapphire window attached to it via thermal paste is placed against the slide with the capillary force from the oil keeping the assembly together with no downward pressure from clamps.

\noindent \textbf{Imaging} Bright-field microscope images and videos are obtained using a Nikon Ti2 microscope with a 10x-magnification, 0.45 NA objective, and a Phantom v9.1 CMOS camera connected to a computer. The condenser is roughly 2/3 closed. The Nikon Perfect Focus System is used to maintain a constant focal plane, which is set to roughly 10~$\mu$m above the bottom of the droplets. Fluorescence images are obtained using a Leica SP8.

\noindent \textbf{Crystallization experiment} For the droplet-based experiments, a temperature protocol is carried out automatically using a programmable temperature controller (TC-720, TE Technology, Inc.). The system is held above the melting temperature for 20 minutes and then dropped in temperature. A custom MATLAB script reads the images as they come in to determine the fraction of droplets that have nucleated crystals over time. If over 90$\%$ of the droplets have a crystal within 1 hour of quenching then the quench will finish and the system will go to a temperature above the melting temperature. Otherwise the temperature is held for a total of 4 hours before remelting. After the system has been melted for 20 minutes, a new quench is performed at a temperature 0.05~$^\circ$C higher than the previous one. This process is continued until fewer than 10$\%$ of the droplets nucleate by the end of 4 hours.

\section*{References}

\bibliography{1-dropletCrystals.bbl}

\begin{thebibliography}{10}
\expandafter\ifx\csname url\endcsname\relax
  \def\url#1{\texttt{#1}}\fi
\expandafter\ifx\csname urlprefix\endcsname\relax\def\urlprefix{URL }\fi
\providecommand{\bibinfo}[2]{#2}
\providecommand{\eprint}[2][]{\url{#2}}

\bibitem{jones2015programmable}
\bibinfo{author}{Jones, M.~R.}, \bibinfo{author}{Seeman, N.~C.} \&
  \bibinfo{author}{Mirkin, C.~A.}
\newblock \bibinfo{title}{Programmable materials and the nature of the {DNA}
  bond}.
\newblock \emph{\bibinfo{journal}{Science}} \textbf{\bibinfo{volume}{347}},
  \bibinfo{pages}{1260901} (\bibinfo{year}{2015}).

\bibitem{Rogers2016}
\bibinfo{author}{Rogers, W.~B.}, \bibinfo{author}{Shih, W.~M.} \&
  \bibinfo{author}{Manoharan, V.~N.}
\newblock \bibinfo{title}{Using {DNA} to program the self-assembly of colloidal
  nanoparticles and microparticles}.
\newblock \emph{\bibinfo{journal}{Nature Reviews Materials}}
  \textbf{\bibinfo{volume}{1}}, \bibinfo{pages}{16008} (\bibinfo{year}{2016}).

\bibitem{laramy2019crystal}
\bibinfo{author}{Laramy, C.~R.}, \bibinfo{author}{O’Brien, M.~N.} \&
  \bibinfo{author}{Mirkin, C.~A.}
\newblock \bibinfo{title}{Crystal engineering with {DNA}}.
\newblock \emph{\bibinfo{journal}{Nature Reviews Materials}}
  \textbf{\bibinfo{volume}{4}}, \bibinfo{pages}{201--224}
  (\bibinfo{year}{2019}).

\bibitem{Mirkin2008Nature}
\bibinfo{author}{Park, S.~Y.} \emph{et~al.}
\newblock \bibinfo{title}{{DNA}-programmable nanoparticle crystallization}.
\newblock \emph{\bibinfo{journal}{Nature}} \textbf{\bibinfo{volume}{451}},
  \bibinfo{pages}{553--556} (\bibinfo{year}{2008}).

\bibitem{Oleg2008Nature}
\bibinfo{author}{Nykypanchuk, D.}, \bibinfo{author}{Maye, M.~M.},
  \bibinfo{author}{van~der Lelie, D.} \& \bibinfo{author}{Gang, O.}
\newblock \bibinfo{title}{{DNA}-guided crystallization of colloidal
  nanoparticles}.
\newblock \emph{\bibinfo{journal}{Nature}} \textbf{\bibinfo{volume}{451}},
  \bibinfo{pages}{549--552} (\bibinfo{year}{2008}).

\bibitem{jones2010dna}
\bibinfo{author}{Jones, M.~R.} \emph{et~al.}
\newblock \bibinfo{title}{{DNA}-nanoparticle superlattices formed from
  anisotropic building blocks}.
\newblock \emph{\bibinfo{journal}{Nature Materials}}
  \textbf{\bibinfo{volume}{9}}, \bibinfo{pages}{913--917}
  (\bibinfo{year}{2010}).

\bibitem{macfarlane2011nanoparticle}
\bibinfo{author}{Macfarlane, R.~J.} \emph{et~al.}
\newblock \bibinfo{title}{Nanoparticle superlattice engineering with {DNA}}.
\newblock \emph{\bibinfo{journal}{Science}} \textbf{\bibinfo{volume}{334}},
  \bibinfo{pages}{204--208} (\bibinfo{year}{2011}).

\bibitem{zhang2013general}
\bibinfo{author}{Zhang, Y.}, \bibinfo{author}{Lu, F.}, \bibinfo{author}{Yager,
  K.~G.}, \bibinfo{author}{Van Der~Lelie, D.} \& \bibinfo{author}{Gang, O.}
\newblock \bibinfo{title}{A general strategy for the {DNA}-mediated
  self-assembly of functional nanoparticles into heterogeneous systems}.
\newblock \emph{\bibinfo{journal}{Nature Nanotechnology}}
  \textbf{\bibinfo{volume}{8}}, \bibinfo{pages}{865} (\bibinfo{year}{2013}).

\bibitem{auyeung2014dna}
\bibinfo{author}{Auyeung, E.} \emph{et~al.}
\newblock \bibinfo{title}{{DNA}-mediated nanoparticle crystallization into
  wulff polyhedra}.
\newblock \emph{\bibinfo{journal}{Nature}} \textbf{\bibinfo{volume}{505}},
  \bibinfo{pages}{73--77} (\bibinfo{year}{2014}).

\bibitem{liu2016diamond}
\bibinfo{author}{Liu, W.} \emph{et~al.}
\newblock \bibinfo{title}{Diamond family of nanoparticle superlattices}.
\newblock \emph{\bibinfo{journal}{Science}} \textbf{\bibinfo{volume}{351}},
  \bibinfo{pages}{582--586} (\bibinfo{year}{2016}).

\bibitem{Crocker2012NComm}
\bibinfo{author}{Casey, M.~T.} \emph{et~al.}
\newblock \bibinfo{title}{Driving diffusionless transformations in colloidal
  crystals using {DNA} handshaking}.
\newblock \emph{\bibinfo{journal}{Nature Communications}}
  \textbf{\bibinfo{volume}{3}}, \bibinfo{pages}{1209} (\bibinfo{year}{2012}).

\bibitem{rogers2015programming}
\bibinfo{author}{Rogers, W.~B.} \& \bibinfo{author}{Manoharan, V.~N.}
\newblock \bibinfo{title}{Programming colloidal phase transitions with {DNA}
  strand displacement}.
\newblock \emph{\bibinfo{journal}{Science}} \textbf{\bibinfo{volume}{347}},
  \bibinfo{pages}{639--642} (\bibinfo{year}{2015}).

\bibitem{Wang2015}
\bibinfo{author}{Wang, Y.} \emph{et~al.}
\newblock \bibinfo{title}{Crystallization of {DNA}-coated colloids}.
\newblock \emph{\bibinfo{journal}{Nature Communications}}
  \textbf{\bibinfo{volume}{6}}, \bibinfo{pages}{7253} (\bibinfo{year}{2015}).

\bibitem{Wang2017NComm}
\bibinfo{author}{Wang, Y.}, \bibinfo{author}{Jenkins, I.~C.},
  \bibinfo{author}{McGinley, J.~T.}, \bibinfo{author}{Sinno, T.} \&
  \bibinfo{author}{Crocker, J.~C.}
\newblock \bibinfo{title}{Colloidal crystals with diamond symmetry at optical
  lengthscales}.
\newblock \emph{\bibinfo{journal}{Nature Communications}}
  \textbf{\bibinfo{volume}{8}}, \bibinfo{pages}{14173} (\bibinfo{year}{2017}).

\bibitem{ducrot2017colloidal}
\bibinfo{author}{Ducrot, {\'E}.}, \bibinfo{author}{He, M.},
  \bibinfo{author}{Yi, G.-R.} \& \bibinfo{author}{Pine, D.~J.}
\newblock \bibinfo{title}{Colloidal alloys with preassembled clusters and
  spheres}.
\newblock \emph{\bibinfo{journal}{Nature materials}}
  \textbf{\bibinfo{volume}{16}}, \bibinfo{pages}{652--657}
  (\bibinfo{year}{2017}).

\bibitem{fang2020two}
\bibinfo{author}{Fang, H.}, \bibinfo{author}{Hagan, M.~F.} \&
  \bibinfo{author}{Rogers, W.~B.}
\newblock \bibinfo{title}{Two-step crystallization and solid--solid transitions
  in binary colloidal mixtures}.
\newblock \emph{\bibinfo{journal}{Proceedings of the National Academy of
  Sciences}} \textbf{\bibinfo{volume}{117}}, \bibinfo{pages}{27927--27933}
  (\bibinfo{year}{2020}).

\bibitem{he2020colloidal}
\bibinfo{author}{He, M.} \emph{et~al.}
\newblock \bibinfo{title}{Colloidal diamond}.
\newblock \emph{\bibinfo{journal}{Nature}} \textbf{\bibinfo{volume}{585}},
  \bibinfo{pages}{524--529} (\bibinfo{year}{2020}).

\bibitem{oxtoby1992homogeneous}
\bibinfo{author}{Oxtoby, D.~W.}
\newblock \bibinfo{title}{Homogeneous nucleation: theory and experiment}.
\newblock \emph{\bibinfo{journal}{Journal of Physics: Condensed Matter}}
  \textbf{\bibinfo{volume}{4}}, \bibinfo{pages}{7627} (\bibinfo{year}{1992}).

\bibitem{Rogers2011PNAS}
\bibinfo{author}{Rogers, W.~B.} \& \bibinfo{author}{Crocker, J.~C.}
\newblock \bibinfo{title}{Direct measurements of {DNA}-mediated colloidal
  interactions and their quantitative modeling}.
\newblock \emph{\bibinfo{journal}{Proceedings of the National Academy of
  Sciences of the United States of America}} \textbf{\bibinfo{volume}{108}},
  \bibinfo{pages}{15687--15692} (\bibinfo{year}{2011}).

\bibitem{Miranda2018SoftMatter}
\bibinfo{author}{Lee-Thorp, J.~P.} \& \bibinfo{author}{Holmes-Cerfon, M.}
\newblock \bibinfo{title}{Modeling the relative dynamics of {DNA}-coated
  colloids}.
\newblock \emph{\bibinfo{journal}{Soft Matter}} \textbf{\bibinfo{volume}{14}},
  \bibinfo{pages}{8147--8159} (\bibinfo{year}{2018}).

\bibitem{Dreyfus2010}
\bibinfo{author}{Dreyfus, R.} \emph{et~al.}
\newblock \bibinfo{title}{Aggregation-disaggregation transition of {DNA}-coated
  colloids: Experiments and theory}.
\newblock \emph{\bibinfo{journal}{Phys. Rev. E}} \textbf{\bibinfo{volume}{81}},
  \bibinfo{pages}{041404} (\bibinfo{year}{2010}).

\bibitem{Michele2013}
\bibinfo{author}{Di~Michele, L.} \& \bibinfo{author}{Eiser, E.}
\newblock \bibinfo{title}{Developments in understanding and controlling self
  assembly of {DNA}-functionalized colloids}.
\newblock \emph{\bibinfo{journal}{Phys. Chem. Chem. Phys.}}
  \textbf{\bibinfo{volume}{15}}, \bibinfo{pages}{3115--3129}
  (\bibinfo{year}{2013}).

\bibitem{ross2015nanoscale}
\bibinfo{author}{Ross, M.~B.}, \bibinfo{author}{Ku, J.~C.},
  \bibinfo{author}{Vaccarezza, V.~M.}, \bibinfo{author}{Schatz, G.~C.} \&
  \bibinfo{author}{Mirkin, C.~A.}
\newblock \bibinfo{title}{Nanoscale form dictates mesoscale function in
  plasmonic {DNA}--nanoparticle superlattices}.
\newblock \emph{\bibinfo{journal}{Nature nanotechnology}}
  \textbf{\bibinfo{volume}{10}}, \bibinfo{pages}{453} (\bibinfo{year}{2015}).

\bibitem{sun2018design}
\bibinfo{author}{Sun, L.}, \bibinfo{author}{Lin, H.},
  \bibinfo{author}{Kohlstedt, K.~L.}, \bibinfo{author}{Schatz, G.~C.} \&
  \bibinfo{author}{Mirkin, C.~A.}
\newblock \bibinfo{title}{Design principles for photonic crystals based on
  plasmonic nanoparticle superlattices}.
\newblock \emph{\bibinfo{journal}{Proceedings of the National Academy of
  Sciences}} \textbf{\bibinfo{volume}{115}}, \bibinfo{pages}{7242--7247}
  (\bibinfo{year}{2018}).

\bibitem{Park2019ACS}
\bibinfo{author}{Park, S.~H.}, \bibinfo{author}{Park, H.},
  \bibinfo{author}{Hur, K.} \& \bibinfo{author}{Lee, S.}
\newblock \bibinfo{title}{Design of {DNA} origami diamond photonic crystals}.
\newblock \emph{\bibinfo{journal}{ACS Applied Bio Materials}}
  \textbf{\bibinfo{volume}{3}}, \bibinfo{pages}{747--756}
  (\bibinfo{year}{2019}).

\bibitem{Pound1952JACS}
\bibinfo{author}{Pound, G.~M.} \& \bibinfo{author}{Mer, V. K.~L.}
\newblock \bibinfo{title}{Kinetics of crystalline nucleus formation in
  supercooled liquid tin}.
\newblock \emph{\bibinfo{journal}{Journal of the American Chemical Society}}
  \textbf{\bibinfo{volume}{74}}, \bibinfo{pages}{2323--2332}
  (\bibinfo{year}{1952}).

\bibitem{galkin1999direct}
\bibinfo{author}{Galkin, O.} \& \bibinfo{author}{Vekilov, P.~G.}
\newblock \bibinfo{title}{Direct determination of the nucleation rates of
  protein crystals}.
\newblock \emph{\bibinfo{journal}{The Journal of Physical Chemistry B}}
  \textbf{\bibinfo{volume}{103}}, \bibinfo{pages}{10965--10971}
  (\bibinfo{year}{1999}).

\bibitem{akella2014emulsion}
\bibinfo{author}{Akella, S.~V.}, \bibinfo{author}{Mowitz, A.},
  \bibinfo{author}{Heymann, M.} \& \bibinfo{author}{Fraden, S.}
\newblock \bibinfo{title}{Emulsion-based technique to measure protein crystal
  nucleation rates of lysozyme}.
\newblock \emph{\bibinfo{journal}{Crystal Growth \& Design}}
  \textbf{\bibinfo{volume}{14}}, \bibinfo{pages}{4487--4509}
  (\bibinfo{year}{2014}).

\bibitem{Sear2014CrystEngCom}
\bibinfo{author}{Sear, R.~P.}
\newblock \bibinfo{title}{Quantitative studies of crystal nucleation at
  constant supersaturation: experimental data and models}.
\newblock \emph{\bibinfo{journal}{CrystEngComm}} \textbf{\bibinfo{volume}{16}},
  \bibinfo{pages}{6506--6522} (\bibinfo{year}{2014}).

\bibitem{Palberg1999CondensedMatter}
\bibinfo{author}{Palberg, T.}
\newblock \bibinfo{title}{Crystallization kinetics of repulsive colloidal
  spheres}.
\newblock \emph{\bibinfo{journal}{Journal of Physics: Condensed Matter}}
  \textbf{\bibinfo{volume}{11}}, \bibinfo{pages}{R323--R360}
  (\bibinfo{year}{1999}).

\bibitem{lowensohn2019linker}
\bibinfo{author}{Lowensohn, J.}, \bibinfo{author}{Oyarz{\'u}n, B.},
  \bibinfo{author}{Paliza, G.~N.}, \bibinfo{author}{Mognetti, B.~M.} \&
  \bibinfo{author}{Rogers, W.~B.}
\newblock \bibinfo{title}{Linker-mediated phase behavior of dna-coated
  colloids}.
\newblock \emph{\bibinfo{journal}{Physical Review X}}
  \textbf{\bibinfo{volume}{9}}, \bibinfo{pages}{041054} (\bibinfo{year}{2019}).

\bibitem{saito1996statistical}
\bibinfo{author}{Saito, Y.}
\newblock \emph{\bibinfo{title}{Statistical physics of crystal growth}}
  (\bibinfo{publisher}{World Scientific}, \bibinfo{year}{1996}).

\bibitem{theBigMan1905}
\bibinfo{author}{Einstein, A.}
\newblock \bibinfo{title}{Über die von der molekularkinetischen theorie der
  wärme geforderte bewegung von in ruhenden flüssigkeiten suspendierten
  teilchen}.
\newblock \emph{\bibinfo{journal}{Annalen der Physik}}
  \textbf{\bibinfo{volume}{322}}, \bibinfo{pages}{549--560}
  (\bibinfo{year}{1905}).

\bibitem{dombrowski2010modeling}
\bibinfo{author}{Dombrowski, R.~D.}, \bibinfo{author}{Litster, J.~D.},
  \bibinfo{author}{Wagner, N.~J.} \& \bibinfo{author}{He, Y.}
\newblock \bibinfo{title}{Modeling the crystallization of proteins and small
  organic molecules in nanoliter drops}.
\newblock \emph{\bibinfo{journal}{AIChE journal}}
  \textbf{\bibinfo{volume}{56}}, \bibinfo{pages}{79--91}
  (\bibinfo{year}{2010}).

\bibitem{heymann2014room}
\bibinfo{author}{Heymann, M.} \emph{et~al.}
\newblock \bibinfo{title}{Room-temperature serial crystallography using a
  kinetically optimized microfluidic device for protein crystallization and
  on-chip x-ray diffraction}.
\newblock \emph{\bibinfo{journal}{IUCrJ}} \textbf{\bibinfo{volume}{1}},
  \bibinfo{pages}{349--360} (\bibinfo{year}{2014}).

\bibitem{Weitz2001Science}
\bibinfo{author}{Gasser, U.}, \bibinfo{author}{Weeks, E.~R.},
  \bibinfo{author}{Schofield, A.}, \bibinfo{author}{Pusey, P.~N.} \&
  \bibinfo{author}{Weitz, D.~A.}
\newblock \bibinfo{title}{Real-space imaging of nucleation and growth in
  colloidal crystallization}.
\newblock \emph{\bibinfo{journal}{Science}} \textbf{\bibinfo{volume}{292}},
  \bibinfo{pages}{258} (\bibinfo{year}{2001}).

\bibitem{phillips2012physical}
\bibinfo{author}{Phillips, R.}, \bibinfo{author}{Kondev, J.},
  \bibinfo{author}{Theriot, J.} \& \bibinfo{author}{Garcia, H.}
\newblock \emph{\bibinfo{title}{Physical biology of the cell}}
  (\bibinfo{publisher}{Garland Science}, \bibinfo{year}{2012}).

\bibitem{Feng2013NatMat}
\bibinfo{author}{Feng, L.} \emph{et~al.}
\newblock \bibinfo{title}{Cinnamate-based {DNA} photolithography}.
\newblock \emph{\bibinfo{journal}{Nature Materials}}
  \textbf{\bibinfo{volume}{12}}, \bibinfo{pages}{747--753}
  (\bibinfo{year}{2013}).

\bibitem{Lee2019}
\bibinfo{author}{Lee, S.}, \bibinfo{author}{Zheng, C.~Y.},
  \bibinfo{author}{Bujold, K.~E.} \& \bibinfo{author}{Mirkin, C.~A.}
\newblock \bibinfo{title}{A cross-linking approach to stabilizing
  stimuli-responsive colloidal crystals engineered with {DNA}}.
\newblock \emph{\bibinfo{journal}{J. Am. Chem. Soc.}}
  \textbf{\bibinfo{volume}{141}}, \bibinfo{pages}{11827--11831}
  (\bibinfo{year}{2019}).

\bibitem{Allahyarov2015NComm}
\bibinfo{author}{Allahyarov, E.}, \bibinfo{author}{Sandomirski, K.},
  \bibinfo{author}{Egelhaaf, S.~U.} \& \bibinfo{author}{L{\"o}wen, H.}
\newblock \bibinfo{title}{Crystallization seeds favour crystallization only
  during initial growth}.
\newblock \emph{\bibinfo{journal}{Nature Communications}}
  \textbf{\bibinfo{volume}{6}}, \bibinfo{pages}{7110} (\bibinfo{year}{2015}).

\bibitem{mirkin1996dna}
\bibinfo{author}{Mirkin, C.~A.}, \bibinfo{author}{Letsinger, R.~L.},
  \bibinfo{author}{Mucic, R.~C.} \& \bibinfo{author}{Storhoff, J.~J.}
\newblock \bibinfo{title}{A {DNA}-based method for rationally assembling
  nanoparticles into macroscopic materials}.
\newblock \emph{\bibinfo{journal}{Nature}} \textbf{\bibinfo{volume}{382}},
  \bibinfo{pages}{607--609} (\bibinfo{year}{1996}).

\bibitem{jacobs2016self}
\bibinfo{author}{Jacobs, W.~M.} \& \bibinfo{author}{Frenkel, D.}
\newblock \bibinfo{title}{Self-assembly of structures with addressable
  complexity}.
\newblock \emph{\bibinfo{journal}{Journal of the American Chemical Society}}
  \textbf{\bibinfo{volume}{138}}, \bibinfo{pages}{2457--2467}
  (\bibinfo{year}{2016}).

\bibitem{Pine2015ChemMater}
\bibinfo{author}{Oh, J.~S.}, \bibinfo{author}{Wang, Y.}, \bibinfo{author}{Pine,
  D.~J.} \& \bibinfo{author}{Yi, G.-R.}
\newblock \bibinfo{title}{High-density {PEO}-b-{DNA} brushes on polymer
  particles for colloidal superstructures}.
\newblock \emph{\bibinfo{journal}{Chem. Mater.}} \textbf{\bibinfo{volume}{27}},
  \bibinfo{pages}{8337--8344} (\bibinfo{year}{2015}).

\end{thebibliography}


\begin{thebibliography}{10}

\bibitem{Pine2015ChemMater}
J.~S. Oh, Y.~Wang, D.~J. Pine, G.-R. Yi, High-density {PEO}-b-{DNA} brushes on
  polymer particles for colloidal superstructures, {\it Chem. Mater.\/} {\bf
  27}, 8337 (2015).

\bibitem{Crocker1996Colloid}
J.~C. Crocker, D.~G. Grier, Methods of digital video microscopy for colloidal
  studies, {\it Journal of Colloid and Interface Science\/} {\bf 179}, 298
  (1996).

\bibitem{wang2015crystallization}
Y.~Wang, Y.~Wang, X.~Zheng, {\'E}.~Ducrot, J.~S. Yodh, M.~Weck, D.~J. Pine,
  Crystallization of {DNA}-coated colloids, {\it Nature Communications\/} {\bf
  6}, 7253 (2015).

\bibitem{charbonneau2007gas}
P.~Charbonneau, D.~Frenkel, Gas-solid coexistence of adhesive spheres, {\it The
  Journal of Chemical Physics\/} {\bf 126}, 196101 (2007).

\bibitem{Rogers2011PNAS}
W.~B. Rogers, J.~C. Crocker, Direct measurements of {DNA}-mediated colloidal
  interactions and their quantitative modeling, {\it Proceedings of the
  National Academy of Sciences of the United States of America\/} {\bf 108},
  15687 (2011).

\bibitem{lowensohn2019linker}
J.~Lowensohn, B.~Oyarz{\'u}n, G.~N. Paliza, B.~M. Mognetti, W.~B. Rogers,
  Linker-mediated phase behavior of dna-coated colloids, {\it Physical Review
  X\/} {\bf 9}, 041054 (2019).

\bibitem{palberg1999crystallization}
T.~Palberg, Crystallization kinetics of repulsive colloidal spheres, {\it
  Journal of Physics: Condensed Matter\/} {\bf 11}, R323 (1999).

\bibitem{Wang2015}
Y.~Wang, Y.~Wang, X.~Zheng, {\'E}.~Ducrot, J.~S. Yodh, M.~Weck, D.~J. Pine,
  Crystallization of {DNA}-coated colloids, {\it Nature Communications\/} {\bf
  6}, 7253 (2015).

\bibitem{Miranda2018SoftMatter}
J.~P. Lee-Thorp, M.~Holmes-Cerfon, Modeling the relative dynamics of
  {DNA}-coated colloids, {\it Soft Matter\/} {\bf 14}, 8147 (2018).

\bibitem{sear2007nucleation}
R.~P. Sear, Nucleation: {T}heory and applications to protein solutions and
  colloidal suspensions, {\it Journal of Physics: Condensed Matter\/} {\bf 19},
  033101 (2007).

\bibitem{oxtoby1992homogeneous}
D.~W. Oxtoby, Homogeneous nucleation: theory and experiment, {\it Journal of
  Physics: Condensed Matter\/} {\bf 4}, 7627 (1992).

\bibitem{libbrecht2003growth}
K.~G. Libbrecht, Growth rates of the principal facets of ice between
  $-10^\circ${C} and $-40^\circ${C}, {\it Journal of Crystal Growth\/} {\bf
  247}, 530 (2003).

\bibitem{saito1996statistical}
Y.~Saito, {\it Statistical physics of crystal growth\/} (World Scientific,
  1996).

\end{thebibliography}

\begin{addendum}
\item[Acknowledgements]We thank Michael F. Hagan, Seth Fraden, Vinothan N. Manoharan, and Rees F. Garmann for helpful discussions; Maria Eleni Moustaka for help in fabricating our microfluidic device and providing access to syringe pumps; and Emily Gehrels for help with the colloid functionalization. This work was supported by the National Science Foundation (DMR-1710112).

\item[Author Contributions] A.H., W.M.J, and W.B.R designed research; A.H. performed research; W. M. J. developed the theory; A.H., W.M.J., and W.B.R analyzed data, and A.H., W.M.J., and W.B.R. wrote the paper.

\item[Competing Interests] The authors declare that they have no competing financial interests.

\item[Correspondence] Correspondence and requests for materials should be addressed to William M. Jacobs~(email: wjacobs@princeton.edu) and W. Benjamin Rogers~(email: wrogers@brandeis.edu).
\end{addendum}

\end{document}


\title{
  \LARGE Supplementary Materials for
  \\
  \bigskip
  \Large Classical nucleation and growth of DNA-programmed colloidal crystallization
}


\author{Alexander Hensley$^{1}$, William M. Jacobs$^{2,*}$ \& W. Benjamin Rogers$^{1,*}$\\
  \\
  \normalsize{$^{1}$ Martin A. Fisher School of Physics, Brandeis University, Waltham, MA USA, 02453}\\
    \normalsize{$^{2}$ Department of Chemistry, Princeton University, Princeton, NJ USA, 08544}\\
  \normalsize{$^\ast$To whom correspondence should be addressed; E-mail:
    wjacobs@princeton.edu, wrogers@brandeis.edu}\\
}

\date{}

\maketitle


%
%
%
%
%

\pagebreak
\part*{Materials and methods}

\section{Synthesizing DNA-coated colloids}
We prepare DNA-grafted colloidal particles using a combination of strain-promoted dibenzocyclooctyne (DBCO) click chemistry and physical grafting, following a modified version of the methods described by Pine and co-workers\cite{Pine2015ChemMater}. In brief, we functionalize polystyrene-block-poly(ethylene oxide) (PS-b-PEO) with an azide group, adsorb the azide-modified PS-b-PEO onto the surface of polystyrene colloidal particles, and then attach DBCO-modified DNA to the PS-b-PEO via click chemistry. We functionalize PS-b-PEO $(M_w=$6500~g/mol and 3800~g/mol, Polymer Source, Inc.) with an azide group. First we attach a mesyl group via methanesulfonyl chloride (471259, Sigma-Aldrich), and then we replace the mesyl group with an azide group $N_3$ via sodium azide (S2002-5G, Sigma-Aldrich). Next, we incorporate the PS-b-PEO-$N_3$ onto the surface of 600-nm-diameter polystrene particles (S37495, Molecular Probes) by swelling the particles with tetrahydrofuran (THF, 99.9\% inhibitor free 401757, Sigma-Aldrich) in an aqueous solution of PS-b-PEO-$N_3$. Next, we add additional deionized (DI) water to the solution and wait for an hour in order to deswell the particles. We then wash the particles five times by repeated centrifugation and resuspension in DI water. Finally, we attach DBCO-modified DNA molecules via a click reaction with the azide groups in a solution of tris-EDTA buffer (pH 8), pluronic F-127 (51181981, Sigma-Aldrich), and sodium chloride. We rotate the reaction mixture end-over-end in an oven at $60^\circ$C for 24 hours and then wash the particles five times in DI water. We store the particles at a concentration of 1\% (v/v) in DI water at $4^\circ$C. The two DNA sequences are 5'-(T)$_{51}$-GAGTTGCGGTAGAC-3' and 5'-(T)$_{51}$-AATGCCTGTCTACC-3'.

\section{Making droplets with microfluidics}
We use PDMS-based microfluidics to prepare colloid-filled droplets. The droplets are stabilized by a fluorinated surfactant (RAN) suspended in fluorinated oil (HFE-7500) at a concentration of 2\%. The droplets contain an aqueous solution of colloidal particles, 5~mM Tris/0.5~mM EDTA buffer, and 500~mM NaCl.\\

\begin{figure}
    \centering
    \includegraphics{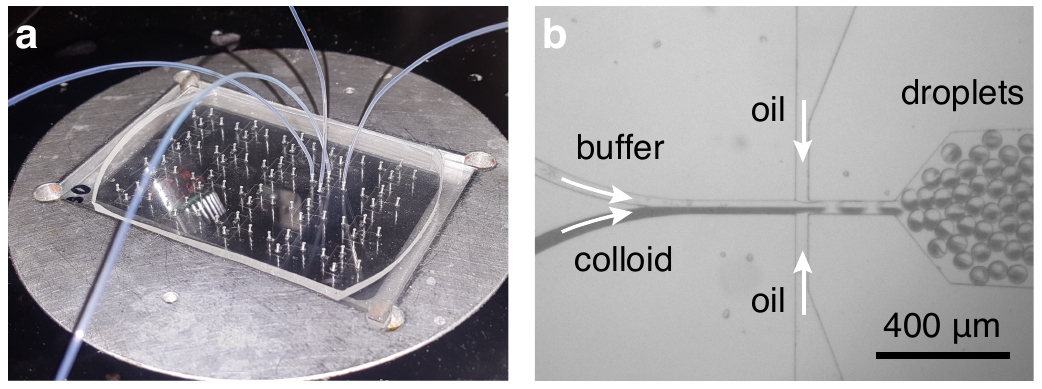}
    \caption{Photographs of the microfluidic chip and the device under operation. (\textbf{a}) A photograph of the droplet-maker on a microscope with connected inlet and outlet tubes. The chip contains an array of 24 identical devices with one in use at any given time. The device has four tubes coming out of it, two are the inlets for the two components of the aqueous phase, one is the inlet for the oil phase, and one is the outlet for the droplets. The inlet tubes lead to three computer controlled syringe pumps and the outlet leads to an Eppendorf tube. (\textbf{b}) A brightfield micrograph of a device in use. The two aqueous inlet streams meet into a single stream before becoming squeezed into droplets by the oil phase. The droplets exit via an output tube. The central channels are roughly 50~$\mu$m in width. The arrows represent the direction of fluid flow.}
    \label{fig:dropmaker}
\end{figure}

\noindent\textbf{Design.} We use a PDMS-based, microfluidic drop maker. Our device is comprised of two inlets for different aqueous phases that combine on-chip, and one inlet for an oil phase which splits into two opposing streams that terminate perpendicular to the combined aqueous stream (Figure~\ref{fig:dropmaker}a). The two aqueous phases are combined at a Y-shaped junction. The combined aqueous phase is then pinched off from two sides by the oil phase to form droplets(Figure~\ref{fig:dropmaker}b). We collect the resulting droplets in an Eppendorf microcentrifuge tube.\\

\noindent \textbf{Manufacture.} We use standard soft-lithography techniques to manufacture our microfluidic drop-makers. We spincoat liquid SU-8 photoresist (SU-8 2075, MicroChem) onto a silicon wafer (3-76-024-V-B, Silicon Materials Inc.) and then place a photomask (Output City) on the wafer. We use a Manual Mask Aligner System (ABM-USA) to make the SU-8 thickness 80~$\mu$m and then we cure the photoresist with UV light. Next we wash away the uncrosslinked photoresist with propylene glycol monomethyl ether acetate (484431, Sigma-Aldrich) to create a negative master of the microfluidic device on the wafer. 

We then create a mold from the master. We line a Petri dish with aluminum foil and place the etched wafer into the dish. We mix polydimethylsiloxane (PDMS) and crosslinker (1673921, Dow Chemical Company) in a 10 to 1 (w/w) ratio with a Thinky AR-250 planetary centrifugal mixer and pour it into the Petri dish. We degas the sample and then place it in an oven at 70 degrees centigrade for at least 2 hours to cure. We separate the cured PDMS from the master with a hobby knife and create inlet and outlet channels with a core sampling tool (69039-07, Electron Microscopy Sciences). Finally, we bond the PDMS device to a glass microscope slide (2947-75X50, Corning) with oxygen-plasma treatment for 45 seconds.\\

\noindent \textbf{Operation.} We use syringe pumps to operate our microfluidic device to produce monodisperse droplets containing colloidal suspension (Figure~\ref{fig:dropmaker}a). First, we make the channels of the microfluidic device hydrophobic by flushing them with Aquapel (B004NFW5EC, Amazon), leaving it for 30 seconds, and then flushing the channels out again with air to remove the Aquapel. We repeat the process with HFE-7500 oil and then air. Finally, we flow oil and aqueous buffer through their respective inlets to ensure the device functions properly. Specifically, we control the flowrates of three 3-ml syringes independently with syringe pumps (98-2662, Harvard Apparatus) connected to the device via tubing (06417-11, Cole-Palmer) that is slightly larger in the diameter than the holes to ensure a snug fit. One syringe feeds in the HFE-7500 (3M) with 2\% RAN fluorosurfactant (008-FluoroSurfactant-5wtH-20G, RAN Biotechnologies). The other two syringes feed in aqueous solutions, which are combined on-chip (Figure~\ref{fig:dropmaker}B). One of the aqueous phases contains 1M NaCl in 1xTE buffer (1~mM EDTA/10~mM Tris), and the other contains either DI water or the DNA-coated particles suspended at twice the desired volume fraction in DI water. We use flow rates of 800~$\mu$l/hr for the oil phase and 400~$\mu$l/hr for each of the aqueous phases to obtain droplets with diameters of roughly 60~$\mu$m.

Due to the small amount of DNA-coated colloids that we produce, we cannot fill the entire syringe with colloidal suspension. Therefore, we use a method to minimize the volume of DNA-coated colloids needed for a given experiment. Specifically, we fill a syringe with HFE-7500, then we place the end of the tube connected to this syringe into an Eppendorf containing the colloidal suspension and use the syringe pump in reverse to pull a 10-$\mu$l-volume plug of colloidal suspension into the tube. We then pull a millimeter-long plug of air into the tube followed by enough 1xTE to fill approximately 0.5~m of tubing. This extra volume of buffer ensures that the droplets are uniform before our colloidal suspension passes through the microfluidic device. Since the particle-filled fluid is separated by a cushion of air on both sides, the particles move through the tubing and the device as a uniform plug and do not spread due to Taylor dispersion. We collect the resulting droplets in an Eppendorf microcentrifuge tube at the exit of the outlet tubing. These droplets remain stable in solution for a period of many months. Though the droplets are stable for months, we use them within a week to minimize any irreversible aggregation that might happen over long periods of time.

\section{Conducting nucleation and growth experiments}


We use optical microscopy to conduct droplet-based measurements of crystal nucleation and growth. The basic workflow includes making a sample chamber, imaging the sample using an inverted optical microscope, controlling the sample's temperature, and then recording time-lapse images as crystals nucleate and grow.\\ 

\noindent \textbf{Making the sample chamber.} We prepare sample chambers comprised of a rectangular capillary filled with colloidal suspension and glued to a glass coverslip. We cut a 100-$\mu$m-thick, 2-mm-wide glass rectangular capillary (5012, VitroCom) to 3~cm in length with a glass scoring pen and hold it in place with a pair of clamping tweezers. We transfer approximately 2--3~$\mu$l of colloidal suspension into the capillary using a micropipette. We fill the rest of the volume of the capillary with HFE-7500 containing 2\% RAN fluorosurfactant. We add HFE-7500 to each side of the capillary such that the bulk of the droplets are in the middle of the capillary during the following steps. We then use a syringe to apply a thin line of UV curing adhesive (NOA 68, Norland), 3~cm in length, onto the center of a plasma cleaned 24~mm~$\times$~60~mm No. 1 glass coverslip (48393-106, VWR). We place the capillary directly onto the line of adhesive and press evenly to enure that the sample is flat against the coverslip. We then seal the ends of the capillary with the same adhesive. We leave the sample to cure under UV light uncovered for 30~s in order to cure the thin film, and then for another 30~min with aluminum foil covering all but the ends of the capillary to seal the ends.\\

\begin{figure}
    \centering
    \includegraphics{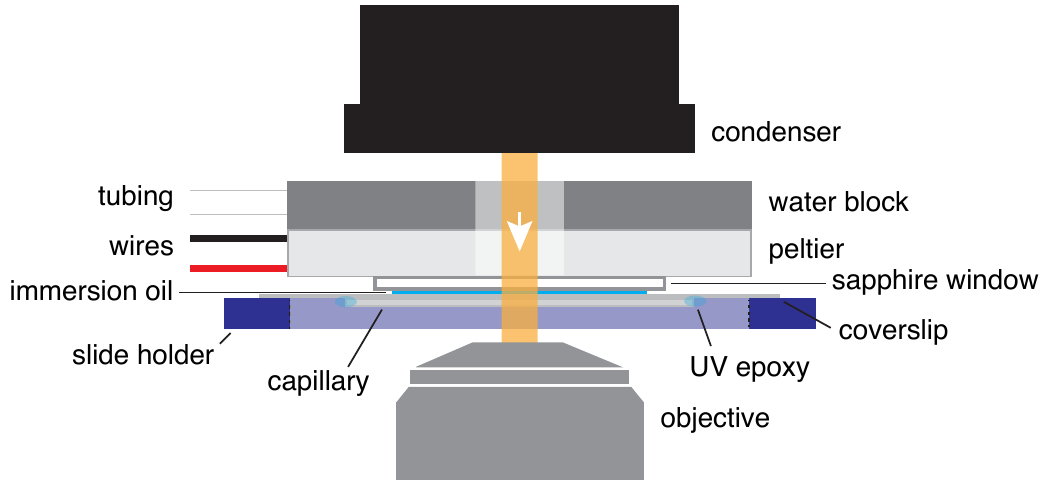}
    \caption{A front facing diagram of the sample chamber and imaging setup.}
    \label{fig:crossSection}
\end{figure}

\noindent \textbf{Imaging the sample.}  We image the sample using a Nikon Eclipse Ti2 inverted microscope equipped with an LED light source (MEE55700, Nikon). We place our sample with the capillary facing down onto an acrylic frame that allows the sample to face the microscope objective without the glue or capillary coming into physical contact with the microscope stage. The frame ensures that the sample remains level and is not bent (Figure~\ref{fig:crossSection}) as other elements are pressed on top of it. We use a 10x-magnification objective (MRD00105, Nikon) to create an image of our sample one the sensor of a CMOS camera (Phantom v9.1).  We focus and center the condenser lens for Koehler illumination and then close the condenser aperture to about $25\%$ to increase the contrast of the edges of the droplets. We use the Nikon Perfect Focus System to keep the sample in focus over the course of our experiment.\\

\noindent \textbf{Acquiring data.} We acquire digital images using a Phantom v9.1 CMOS camera controlled by the PCC 2.8 software. All images have a pixel resolution of $1632\times1200$ and a pixel size of 1.15~$\mu$m. The camera takes $14$-bit grayscale images that are saved as $16$-bit TIFF files in batches of 23 images at a time. We acquire images with a frame rate of 0.1~Hz.\\

\noindent \textbf{Controlling the temperature.} We control the temperature of our sample \emph{in situ} using our temperature-controlled sample chamber and a closed-loop temperature controller. Our custom heating device is illustrated in Figure~\ref{fig:crossSection}. Here we list the components of the device starting from the sample and moving up toward the condenser lens. On the top of the sample we place a droplet of immersion oil (MXA22166, Nikon) to create thermal contact with a sapphire window, which is connected via thermal paste (1446622, Dow Corning) to a Peltier thermoelectric cooler (TEC). The TEC module (CH-109-1.4-1.5, TE-Technology) has a central hole that allows us to image our sample in transmission. A metal waterblock placed on top of the TEC removes waste heat produced during active cooling. Capillary forces due to the thin layer of immersion oil between the sapphire window and coverslip hold the assembly together. 

A temperature controller monitors the sample temperature using a temperature-sensitive resistor attached to the surface of the Peltier heater. The control loop adjusts the current running through the Peltier heater to achieve a user-specified setpoint temperature. We estimate the fluctuations in the sample temperature to be $\pm0.02^\circ C$.

We run a specific temperature protocol to collect nucleation-rate and crystal-size information in a consistent manner. We begin by determining a temperature at which the DNA-grafted particles in the droplets disaggregate, also referred to as melting. We then lower the temperature, or quench, to such a degree that spontaneous aggregation occurs. If spontaneous aggregation does not occur at the chosen temperature we melt the sample and try a lower temperature. We then begin the experiment by melting the sample and quenching it to the temperature at which we found spontaneous aggregation. After each quench, we remelt the sample for 20 minutes and then quench to a temperature that is $0.05^\circ C$ higher than the previous quench. We repeat this process until there is no observable assembly over the full duration of the experiment. Each quench is held for one hour. If less than 90\% of the droplets have formed a crystal after one hour, the quench is held for total duration of $4~hr$.\\

\begin{figure}
    \includegraphics{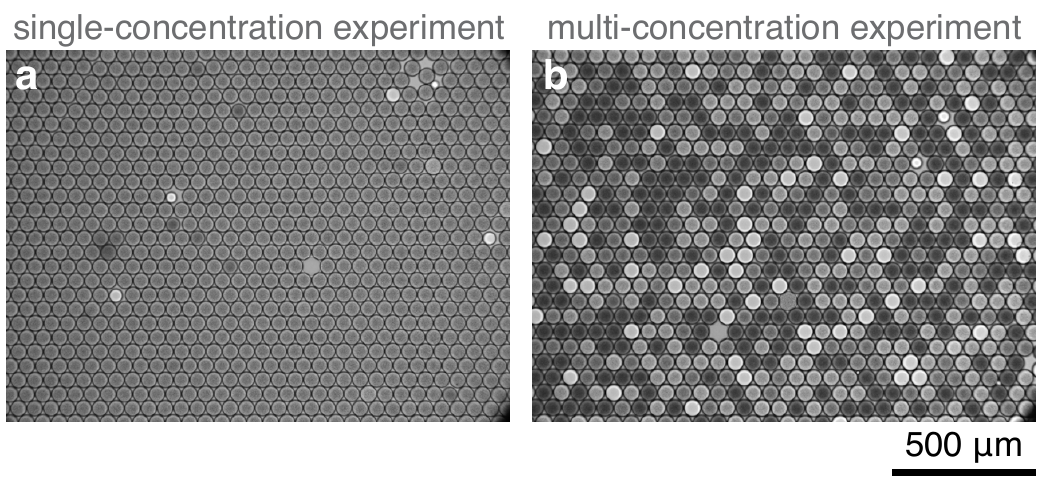}
    \centering
    \caption{Micrographs of the two kinds of experiments we perform.  (\textbf{a}) An example image of a single-concentration experiment of droplets with a particle volume fraction of $0.50\%$. (\textbf{b}) An example image of a multi-concentration experiment with roughly an equal number of droplets of $1.00\%$, $0.50\%$, and $0.25\%$ colloids (v/v), as well as a small number of 0\% droplets. The droplets with lower volume fractions appear  brighter than those with higher concentration. There are roughly $600$ droplets in each image. The temperature in both cases is held above the colloid melting temperature.}
    \label{fig:multiConcImages}
\end{figure}

\noindent \textbf{Running the experiment} We perform two types of droplet-based experiments: one in which all droplets have the same colloid concentration and another in which we image four different droplet types simultaneously. The single-concentration experiment enables us to image roughly 600 nearly identical droplets simultaneously (Figure~\ref{fig:multiConcImages}a). Imaging 600 droplets at once provides good statistical power. However, using droplets of a single concentration has two drawbacks: 1) it is difficult to compare experiments performed at different colloid concentrations due to the finite accuracy of our temperature controller; and 2) it is difficult to calibrate variations in the transmitted light intensity that come from drift in the illumination. Therefore, we perform a second type of experiment that uses droplets of four different concentrations simultaneously to remedy these two issues (Figure~\ref{fig:multiConcImages}b). The droplets in the multi-concentration experiment contain four different colloid concentrations: 0\%, 0.25\%, 0.50\%, and 1.00\% (v/v). We add roughly equal proportions of the 0.25\%, 0.5\%, and 1\% droplets, as well as a small number of colloid-free droplets. The primary advantage of the multi-concentration experiment is that it allows us to make direct comparisons between different concentrations,  ensuring that the entire system is at the same temperature. We use the empty droplets for calibration as they should not change in intensity over time. Thus changes in intensity in the empty droplets come from systemic changes to our imaging setup and can be used to correct the intensities of the other droplets. The disadvantage of this experiment compared to the single-concentration experiment is that the statistical power for each droplet concentration cut by a factor of three. However, we compared the measured nucleation rates for a single-concentration experiment with $0.50\%$ colloids to a multiconcentration-experiment and did not find any noticeable differences.

\section{Running temperature-ramp experiments}
We perform all temperature-ramp experiments in a PCR thermal cycler. In brief, we prepare an emulsion of colloid-filled droplets in 200-microliter-volume PCR tubes, as decsribed above. We place the PCR tubes in a thermal cycler (Bio-Rad C1000 Touch Thermal Cycler) and program a linear temperature ramp, which begins at an initial temperature that is above the colloid melting temperature. We also place a thermistor, which is potted in a dab of thermal compound, in a PCR tube in a neighboring well to monitor the temperature throughout the annealing protocol. The C1000 Touch thermal cycler has a minimum temperature step size of 0.1$^\circ$C. Therefore, we specify the average ramp rate by setting the duration of each step in the protocol. More specifically, we vary the step length from 8~hr to 30~min, yielding average ramp rates that vary from 0.0125$^\circ$C/hr to 0.2$^\circ$C/hr. To maintain as uniform temperature as possible in our emulsion and to minimize evaporation, we set the lid temperature of the thermal cycler to 55$^\circ$C.

After annealing, we measure the probability of forming a single crystal in each drop using optical microscopy. We prepare a sample chamber of crystal-filled droplets, as described above, and image them on a Nikon Eclipse Ti2 inverted microscope. Next we acquire images of 2--3 fields of view, which contain an unbiased sample of roughly 1000 droplets. Finally, we manually inspect each droplet and count the number of distinct crystal domains. Any anomalously large droplets, which we attribute to coalescence sometime during the experiment, are excluded from the analysis. 

\section{Image and data analysis}
\label{sec:imageDataAnalysis}
We extract quantitative information on each of the droplets in the image sequences using image-analysis and data-analysis routines written in MATLAB.\\

\begin{figure}
    \centering
    \includegraphics{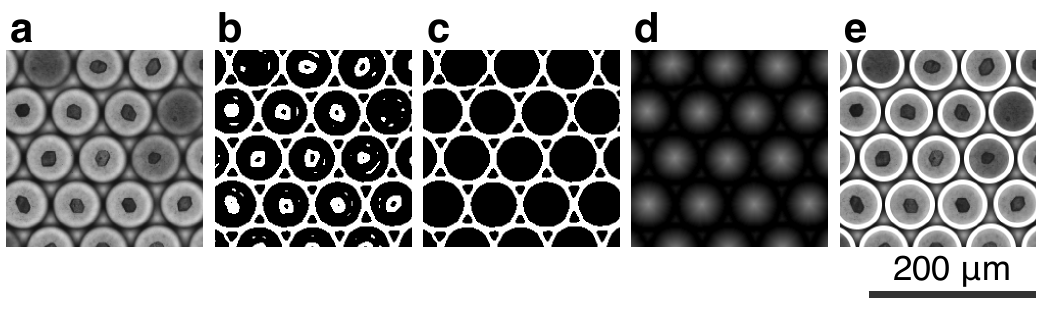}
    \caption{The image transformations done to identify droplet locations efficiently. (\textbf{a}) The raw image. (\textbf{b}) We apply a gaussian blur to the image to dampen high frequency noise, then apply a Laplacian transform to accentuate the droplet edges, and then apply a threshold to binarize the image. (\textbf{c}) We remove all small contiguous features leaving only the connected mesh of droplet edges. (\textbf{d}) We calculate a distance map showing the distance each pixel is to the nearest droplet edge pixel, where the brightest pixels are at the droplet centers. (\textbf{e}) We use a centroiding routine to identify the droplet locations and radii.}
    \label{fig:dropfinding}
\end{figure}

\noindent \textbf{Droplet identification and tracking.} We find the droplet positions and radii for each image using a sequence of image transforms. First we apply a Gaussian filter to the raw images (Figure~\ref{fig:dropfinding}a) to smooth out the high frequency noise. Then we perform a Laplacian convolution to accentuate the edges of the droplets and threshold the image to remove low-intensity noise and create a binary image (Figure~\ref{fig:dropfinding}b). Next we remove any small disconnected features within the droplets (Figure~\ref{fig:dropfinding}c). We perform a Euclidean distance transform to convert the binary image into a grayscale image with a bright spot in the center of every droplet  (Figure~\ref{fig:dropfinding}d). The central pixel value is roughly equal to the radius of the droplet in units of pixels. Lastly, we use a centroiding routine to determine the droplet centers and radii\cite{Crocker1996Colloid} (Figure~\ref{fig:dropfinding}e). For each image, we create a matrix that contains the $x$ and $y$ positions, as well as the droplet radii. Finally, we track the droplets as a function of time using the particle-tracking algorithm developed by Crocker and co-workers \cite{Crocker1996Colloid}. \\

\begin{figure}[H]
    \centering
    \includegraphics{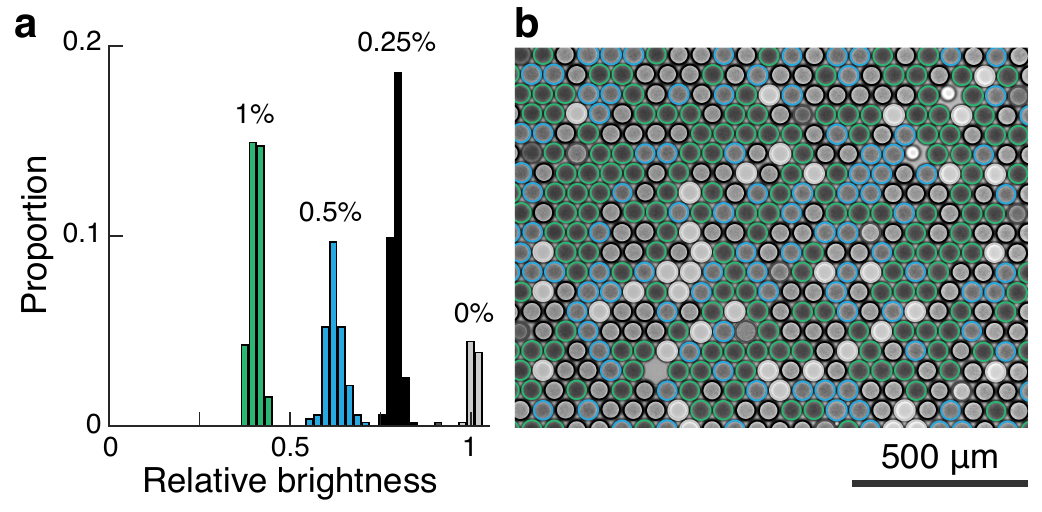}
    \caption{Identification of droplet concentrations from their relative brightness. We bin four different populations of droplets by their transmitted intensities. (\textbf{a}) A histogram of the transmitted intensities of droplets relative to the intensity transmitted through empty droplets. The data is multimodal with each separate cluster in the histogram representing a different concentration. Dark gray bars represent droplets that are ambiguous and are therefore omitted from our analysis. (\textbf{b}) An image from the experiment in which the droplets are highlighted according to their concentration by the colors in (\textbf{a}).}
    \label{fig:multiConcHist}
\end{figure}

\noindent \textbf{Separating the droplets by concentration.} We identify droplets of different concentration by their transmitted light intensities. Figure~\ref{fig:multiConcHist}a shows a histogram of the relative droplet intensities for a sample containing four different droplet concentrations: $0.00\%, 0.25\%, 0.50\%, 1.00\%$. Droplets with a higher concentration appear darker. We separate the populations by applying thresholds between the four different peaks. Figure~\ref{fig:multiConcHist}b shows the droplets color coded according to their inferred concentrations from the histogram. Visual inspection shows that all droplets are classified correctly.

\noindent \textbf{Crystal identification.} We identify crystals within the droplets using a contrast based method. We measure the intensity of each droplet $I(t)$ by averaging the pixel values within an annulus of a user-defined inner and outer radius around the droplet center. We use an annulus instead of the entire droplet to ensure a uniform path length. To determine if a crystal is present we identify all pixels in each droplet that have an intensity that is less than 0.65 times the mean annulus intensity. If there is a contiguous block of these dark pixels greater than a threshold size of 15 pixels, we identify this droplet as containing a crystal. This information is appended to the position and radius matrix for each droplet.\\

\noindent \textbf{Identifying nucleation events.} We determine the time at which crystals first nucleate using a rolling analysis of 100 sequential frames for each droplet, starting at frame one. If a crystal is identified in at least 65 of the 100 frames, we define the first instance of the crystal as the nucleation time. If fewer than 65 of the 100 frames contains a crystal, we slide the 100-frame window forward by one frame and repeat the analysis. We use the sliding window to minimize errors due to any false positives.
\\

\begin{figure}[H]
    \centering
    \includegraphics{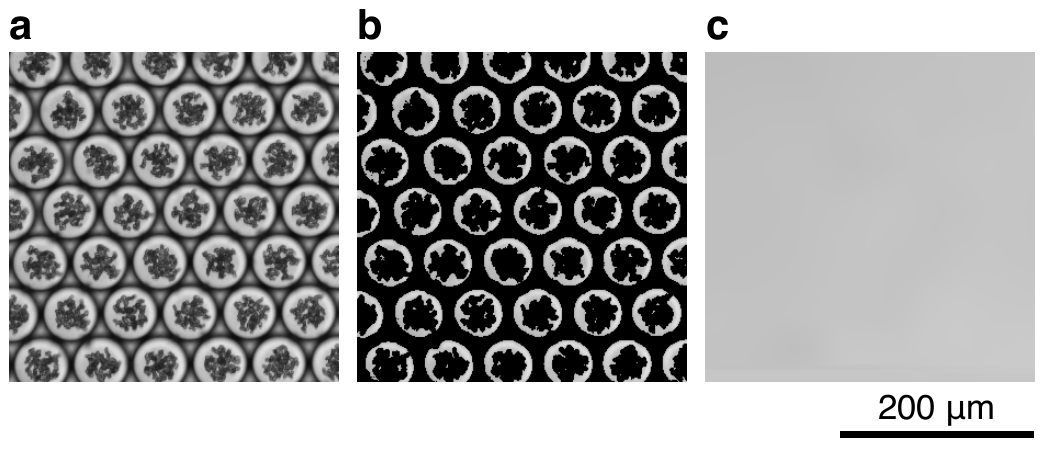}
    \caption{Determining the reference intensity. (\textbf{a}) We acquire a calibration image in which all of the particles are aggregated in the center of the droplets. We threshold the image to remove the aggregated particles (\textbf{b}) and then average the intensity of the remaining pixels. (\textbf{c}) We interpolate the average transmitted intensities from (\textbf{b}) to create a two-dimensional map of the reference intensity $I_0(x,y)$ at every location on the image.}
    \label{fig:calibration}
\end{figure}

\noindent \textbf{Quantifying growth.} We analyze the growth of crystals by comparing the transmitted light intensity, $I(t)$, of each droplet over time to the transmitted intensity of an empty droplet $I_0$. We first create a calibration image to obtain a reference intensity $I_0$ at each point in the image plane. Briefly, we take an image of the system at a low temperature at which the particles are completely aggregated, leaving zero free particles in the bulk (Figure~\ref{fig:calibration}a). We then mask out the aggregates using a threshold (Figure~\ref{fig:calibration}b), and compute the average intensity of each droplet within the user-specified annulus. The intensity of the annulus should be that of an empty droplet, as there are no particles in the fluid phase in this calibration image. Next we construct two-dimensional map of $I_0(x,y)$ at each pixel using two-dimensional interpolation (Figure~\ref{fig:calibration}c). We use this calibration image to compute the relative intensity $I(t)/I_0$ of every droplet in every frame.

We correct for any temporal variations in the reference intensity using the transmitted intensity of the empty droplets. Over the full duration of our experiments, which can take multiple days to complete, the illumination can change with time, either due to small changes in the focus or variations in the lamp intensity. We take advantage of the fact that the relative intensity $I(t)/I_0$ of the empty droplets should not vary with time to correct for any intensity drift. Specifically, we divide the relative intensity $I(t)/I_0$ of each droplet by the value $I(t)/I_0$ averaged over all the empty droplets in each frame. 

\begin{figure}
    \centering
    \includegraphics[]{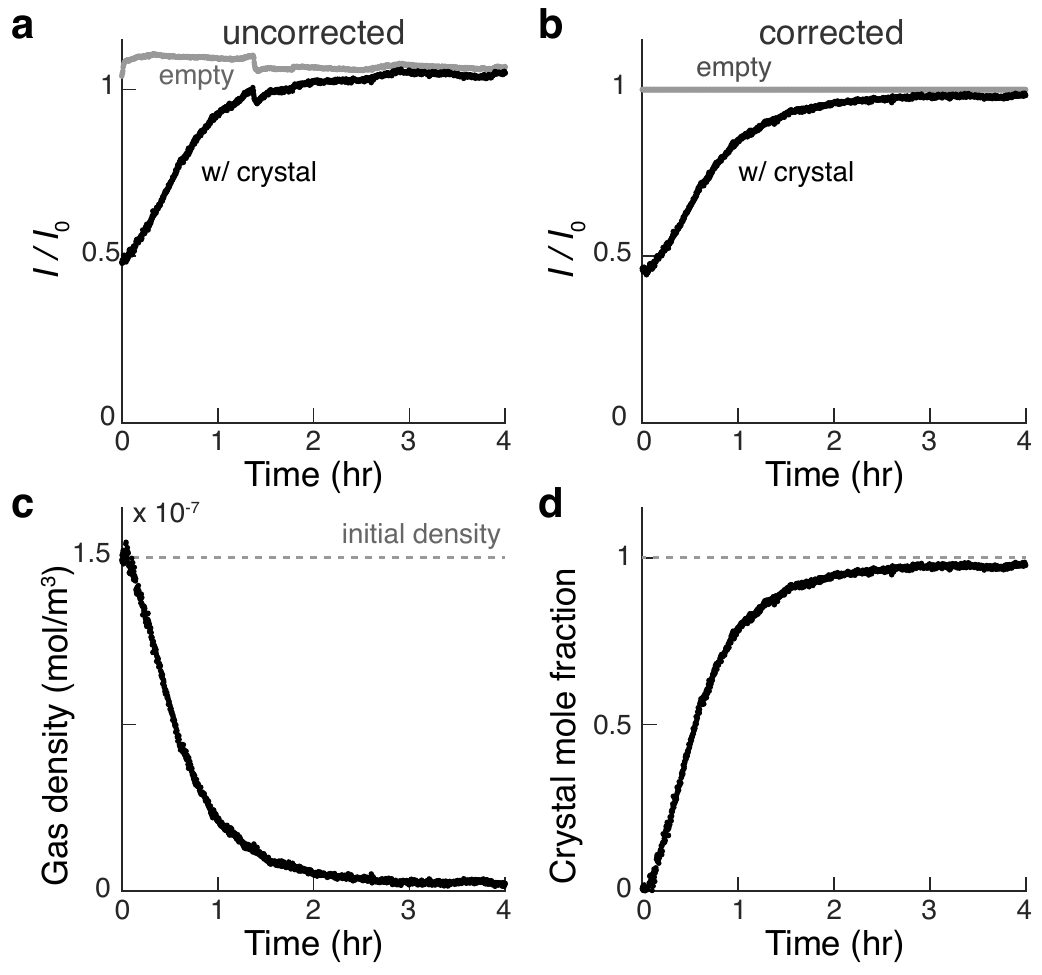}
    \caption{Inferring the gas density and crystal mole fraction from the transmitted light intensity. (\textbf{a}) The relative transmitted light intensity $I(t)/I_0$ as a function of time for an empty droplet and a droplet containing a growing crystal. Variations in $I/I_0$ for the empty droplet are artifacts from focus drift or fluctuations in the lamp intensity. The same variations are seen in $I/I_0$ for the droplet with a crystal. The monotonic increase in $I(t)/I_0$ for the droplet with a crystal is due to the depletion of monomers in the gas phase from crystal growth. (\textbf{b}) Artifacts in $I/I_0$ are corrected by dividing the relative intensity by the relative intensity transmitted through the empty droplets. (\textbf{c}) We infer the gas density as a function of time from the corrected relative intensities using Equation~\ref{eqn:Beer}. (\textbf{d}) We calculate the crystal mole fraction $\chi$ from the gas density in (\textbf{c}) according to $\chi = (\rho_0-\rho(t))/\rho_0$.}
    \label{fig:normalization}
\end{figure}

Figure~\ref{fig:normalization} shows an example of an uncorrected and a corrected trajectory. The relative intensity of the droplet increases as the crystal grows, depleting monomers from the bulk. There are clearly spurious features in the trajectory. When we compare the relative intensity of the droplet to the average behavior of the empty droplets, we find the same spurious features (Figure~\ref{fig:normalization}a). Because the empty droplets should always have a relative intensity of 1, and any deviations from that intensity value are systemic to all the droplets, we can be correct the relative intensities of the droplets by normalizing them with respect to the empty droplets (Figure~\ref{fig:normalization}b).

We use the ratio of the transmitted intensity to the reference intensity to calculate the gas density within each droplet as a function of time. We assume single scattering and infer the number density of the gas phase $\rho(t)$ from the relative intensity $I(t)/I_0$:
\begin{equation}
    \frac{I(t)}{I_0}=\exp{\left[-\rho(t)\sigma L\right]},
    \label{eqn:Beer}
\end{equation}
where $\sigma$ is the scattering cross section, and $L$ is the path length. Although we do not know $\sigma L$ \emph{a priori}, we can infer the value of the product $\sigma L$ from the transmitted light intensities at time $t=0$, when all of the particles are known to be in the gas phase. Then we use the fact that the quantity $\sigma L$ is a constant in time to then infer the number density of the gas phase $\rho(t)$ within each droplet for all subsequent frames (Figure~\ref{fig:normalization}c). 

\begin{figure}
    \centering
    \includegraphics[]{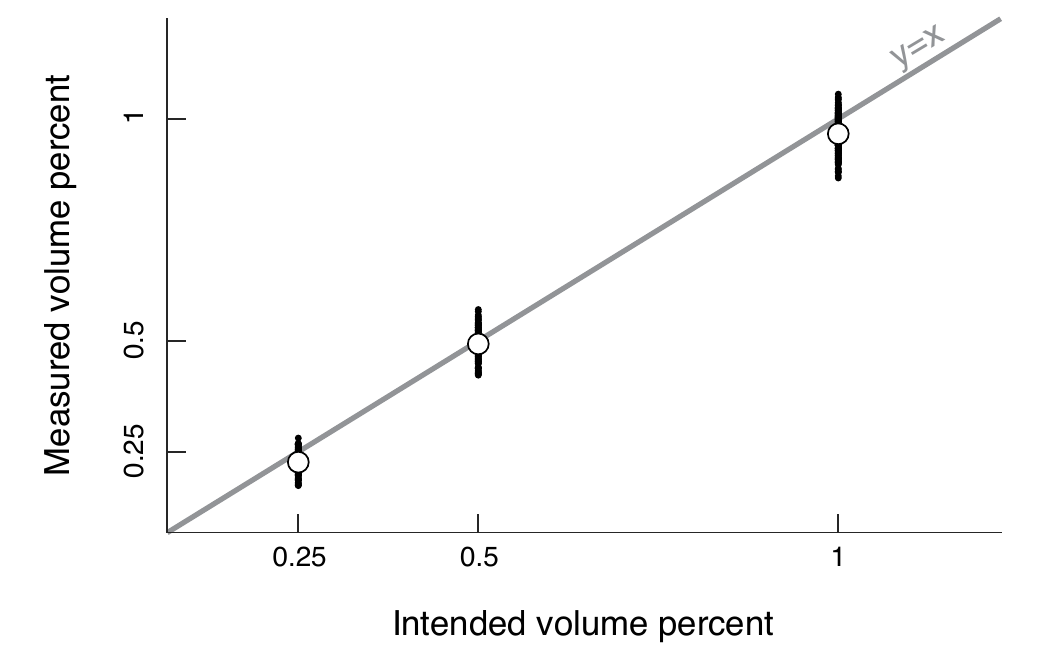}
    \caption{A single-scattering model accurately captures the scattered light intensities. After fitting a single value of $\sigma L$, the product of the scattering cross section $\sigma$ and the pathlength $L$, we measure the volume fractions of three populations of droplets with volume fractions of 0.25\%, 0.5\%, and 1\%. Black points show the measured volume percent for individual droplets; white points show the average volume percentage of roughly 200 droplets of a given concentration. The gray line shows $y=x$, which corresponds to perfect agreement between our experiments and our single-scattering model. The small deviations that we see in the mean volume fraction likely result from small errors in pipetting or small imbalances in the flowrates in our microfluidic device. }
    \label{fig:scattering}
\end{figure}

We confirm that the transmitted light intensity is governed by single scattering and obeys Equation~\ref{eqn:Beer}. Specifically, we fit a single value of $\sigma L$ at $t=0$ to a set of roughly 600 droplets with three particle volume concentrations. We then use this value to infer the colloid volume fraction in each drop individually (Figure~\ref{fig:scattering}). We find that the mean values of the measured volume fractions match the intended volume fractions that we prepared. Furthermore, our measurements also provide an upper bound on the droplet-to-droplet variations in the concentration, which vary by roughly $\pm 6\%$. This value is most likely an upper bound since it includes contributions due to small spatial variations in the lamp intensity in addition to variations in the concentration from droplet to droplet.  \\

Finally, assuming a single crystal forms within each drop, we infer the size of the crystal within each as a function of time from the gas density. Specifically, we determine the number of particles in the crystal $N$ from $N(t) = [\rho_0 - \rho(t)]V_\textrm{drop}$, where $\rho_0$ is the initial number density of the gas phase at $t=0$ and  $V_\textrm{drop}$ is the volume of a droplet. We also calculate the mole fraction of the crystal phase $\chi(t)$ from $\chi(t)=[\rho_0 - \rho(t)]/\rho_0$. Figure~\ref{fig:normalization}d shows the inferred crystal mole fraction within a single drop as a function of time. \\

\begin{figure}[h!]
    \centering
    \includegraphics[]{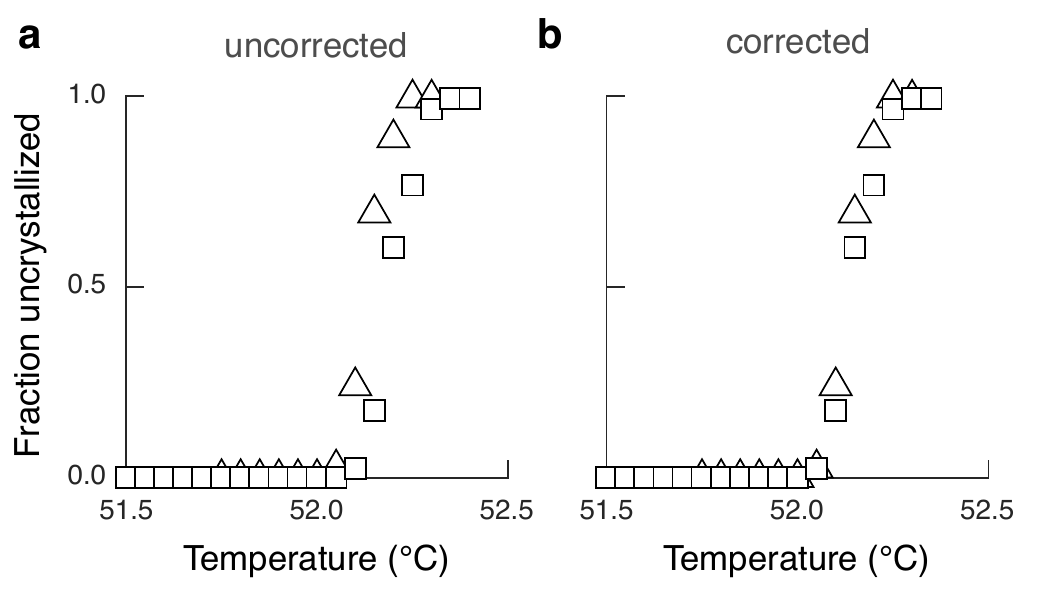}
    \caption{We correct for small systematic errors in the absolute temperature from run to run. More specifically, we shift the temperatures of a given run by a constant offset to align the fraction of uncrystallized droplets that remain after four hours. We attribute the systematic error in the absolute temperature, which is typically around 0.05~$^\circ$C, to small differences in the thermal contact between the thermoelectric cooler and the sample (Fig.~\ref{fig:crossSection}). (\textbf{a}) shows the fraction of droplets that have not yet nucleated after four hours for two different runs of our experiment before any correction. Different shaped symbols correspond to different runs of the experiment. (\textbf{b}) shows the fraction of droplets that have not yet nucleated after four hours post correction. The squares have been shifted by $-$0.05~$^\circ$C.}
    \label{fig:TOffset}
\end{figure}

\noindent \textbf{Correcting for temperature offsets.} Although our temperature controlled sample chamber has a spatial uniformity and precision of roughly $0.01~^\circ$C, we find that the accuracy of the absolute temperature can vary by 0.1--$0.2~^\circ$C from run to run. We attribute these shifts in the absolute temperature to slight differences in the thermal contact between the various layers of our sample chamber (Fig.~\ref{fig:crossSection}). We correct for systematic errors in the absolute temperature by shifting the temperatures to align our measurements of the fraction of droplets that fail to nucleate at the end of a given quench (Figure~\ref{fig:TOffset}). For example, Figure~\ref{fig:TOffset}a shows the fraction of 0.25\% volume fraction droplets that fail to nucleate by the end of a quench as a function of the temperature for two different runs. As the temperature increases, a smaller number of droplets nucleate a crystal within the finite duration of the quench. We see the same trend for both runs, but one run is shifted to higher temperature with respect to the other by $0.07~^\circ$C. We attribute this shift to a systematic error in the measurement of the absolute temperature. Therefore, we offset the temperatures to align the fraction uncrystallized as a function of temperature for each independent run (Figure~\ref{fig:TOffset}b).  \\

\noindent \textbf{Measuring the equilibrium gas density.} We infer the gas density at coexistence from our measurements of the transmitted light intensity. Figure~\ref{fig:equilibriumConc}a shows example snapshots of equilibrated crystals at a variety of temperatures. For each initial colloid concentration and temperature, we fit Equation~\ref{eq:diffusion-limited} to the growth trajectories from individual droplets, taking $\rho_{eq}$, $D$, and $t_\textrm{offset}$ as free parameters. We then average the gas densities at coexistence over all droplets at each temperature, and take this value to be the gas density at equilibrium. We do not notice any systematic dependence of the equilibrium gas density on the initial colloid concentration: All three initial concentrations have roughly the same gas density at coexistence at the same temperature.

Our measurements show that the gas density at coexistence increases exponentially with increasing temperature. Figure~\ref{fig:equilibriumConc}b shows measurements of the equilibrium gas density as a function of temperature. Over a range of roughly 0.25~degrees Celsius, we see that the equilibrium gas density increases by roughly an order of magnitude upon increasing temperature. Furthermore, when plotted on semilog axes, we see that the gas density increases roughly linearly between 52--52.25~degrees Celsius. We note that the equilibrium density continues to decrease at even lower temperatures. However, those low concentrations are below the detection limit of our experimental approach. We determine our detection limit by calculating the standard deviation of the measured bulk concentration of droplets for deep quenches where nearly all of the particles are contained in a dense cluster. For each droplet in the deepest quench of a $1.00\%$ run, we calculated the standard deviation of the bulk concentration for the final 100 frames of the quench and took the mean of all these standard deviations as our minimum detection threshold. We calculate a threshold bulk concentration of $10^{-9}$ mol/m$^3$ and do not consider concentrations at or below this threshold when we perform our fit in Figure~\ref{fig:equilibriumConc}b. For the $32\mu$m in radius droplets we use, this concentration corresponds to around 80 particles.\\

\section{Modeling nucleation and growth}

\begin{table}
  \centering
  \begin{tabular}{c|c|c}
    \textbf{Parameter} & \textbf{Estimated value} & \textbf{Description} \\
    \hline \hline
    $d$ & $6\times10^{-7}\,\mathrm{m}$ & colloid diameter \\
    $\delta$ & $1.5\times10^{-8}\,\mathrm{m}$ & colloid--colloid interaction range \\
    $\phi$ & $0.74$ & crystal packing fraction \\
    $D$ & $1\times10^{-12}\,\mathrm{m}^2\mathrm{s}^{-1}$ & gas-phase diffusion constant \\
    $\kappa$ & $0.1\,\mathrm{s}^{-1}$~\cite{wang2015crystallization} & rolling diffusion rate \\
    $V_{\mathrm{drop}}$ & $1\times10^{-13}\,\mathrm{m}^3$ & droplet volume
  \end{tabular}
  \caption{Descriptions and estimated values of parameters used in the DNA-grafted colloid crystallization model. \label{tab:parameters}}
\end{table}

\subsection{Equilibrium thermodynamics}
\label{sec:thermodynamics}

\begin{figure}[t]
    \centering
    \includegraphics{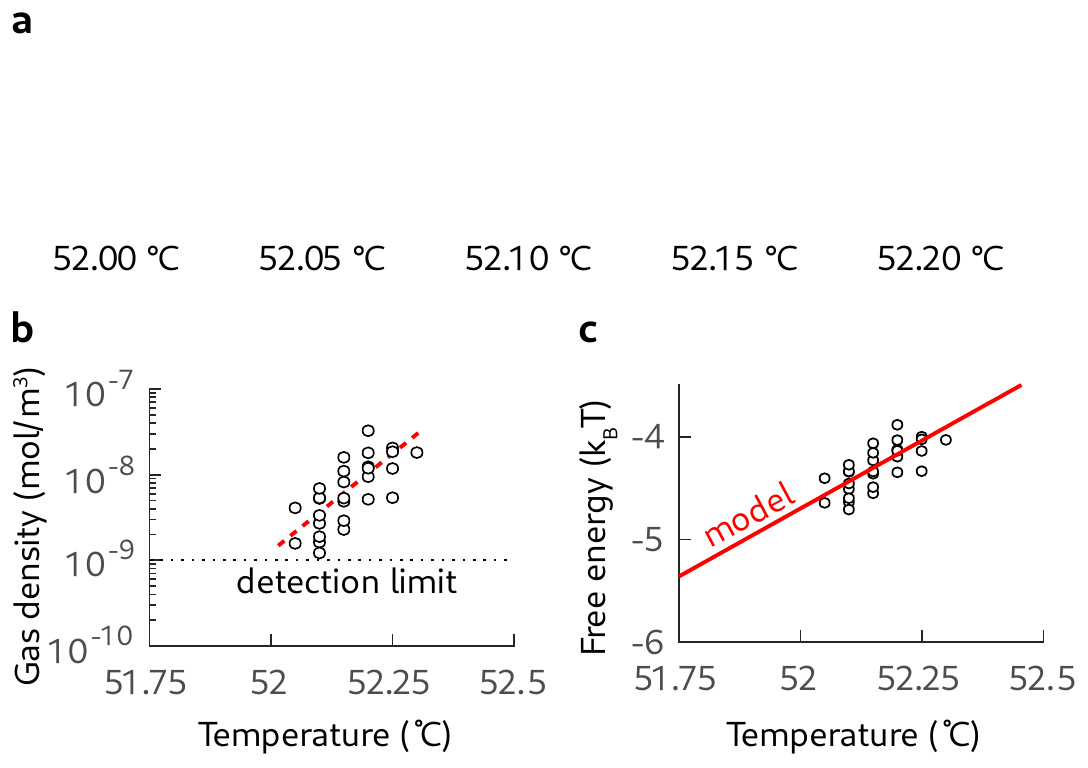}
    \caption{The equilibrium bulk concentration is recorded at each temperature and fit to a line to infer the supersaturation for all experiments. (\textbf{a}) Sample crystals grown to their fullest extent for a variety of temperatures show that as the temperature is increased, the larger equilibrium bulk concentration makes the droplet bulk appear darker. (\textbf{b}) The equilibrium concentration with respect to temperature. We only plot equilibrium concentrations in excess of $10^9$ mol/m$^{-3}$, as that is our detection limit below which the noise in our analysis is greater than the value of the concentration.  A linear fit is indicated by the dashed red line. (\textbf{c}) The estimated binding free energy between a pair of colloidal particles using the model presented in \eqref{eq:epsilon}.}
    \label{fig:equilibriumConc}
\end{figure}

Under the conditions investigated in the experiments, the colloid gas phase can be treated as an ideal gas, since the mean free path in the gas phase is approximately ten times the particle radius.
The equilibrium gas density, $\rho_{\mathrm{eq}}$, in coexistence with the crystal phase is fit to the equation (\figref{fig:equilibriumConc}a,b)
\begin{equation}
  \label{eq:rhoeq}
  \log \left[\rho_{\mathrm{eq}}(T) / \mathrm{M} \right] = 10.5 (T / ^\circ \mathrm{C}) - 573.4.
\end{equation}
The measured gas-phase density is twice the density of each of the two distinct colloidal species.
From these measurements, we define the temperature-dependent supersaturation, $S$,
\begin{equation}
  S(T) = \frac{\rho}{\rho_{\mathrm{eq}}(T)},
\end{equation}
which is the same for each of the colloidal species in the equimolar mixture.

We model the DNA-grafted colloids using the hard-sphere square-well model with colloid diameter $d$ and square-well width $\delta$ (see \tabref{tab:parameters}).
The coordination number in the crystal phase is $z = 8$, which is the number of `unlike' contacts between the two particle species in the CuAu-FCC crystal structure.
Using the cell model presented in Reference~\cite{charbonneau2007gas}, we estimate the binding free energy per colloid--colloid contact, $\epsilon(T)$, according to (\figref{fig:equilibriumConc}c)
\begin{equation}
  \label{eq:epsilon}
  \frac{\epsilon(T)}{k_{\mathrm{B}} T} = \frac{2\left\{\log\left[\rho_{\mathrm{eq}}(T) \delta^3\right]+1\right\}}{z}.
\end{equation}
We thus expect the binding free energy $|\epsilon|$ to be roughly 4 to 5 $k_{\mathrm{B}}T$ over the range of temperatures used in the experiments.
This range of binding free energies is consistent with previous measurements of the colloid--colloid binding free energy~\cite{Rogers2011PNAS,lowensohn2019linker}.
Nonetheless, we note that the absolute value of the binding free energy predicted by the cell model, which influences our prediction of the crystal--vapor surface tension, $\gamma(T)$ (see \secref{sec:surface-tension}), is sensitive to our estimate of the parameter $\delta$.
Consequently, we focus on the temperature dependence of $\epsilon$ and $\gamma$, which is independent of $\delta$, when comparing the cell model to our experimental measurements.

\subsection{Attachment kinetics}

We model the attachment of a free colloidal particle to a crystalline nucleus via a two-step process: First, the particle adsorbs onto the surface of the cluster, and second, the particle diffuses along the surface to a stable position within the crystalline lattice.
Under mild supersaturation, attachment of a particle to a cluster consisting of $i$ colloids can be modeled as an activated process involving an unstable intermediate, $(i+1)^\dagger$:
\begin{center}\includegraphics[width=0.35\textwidth]{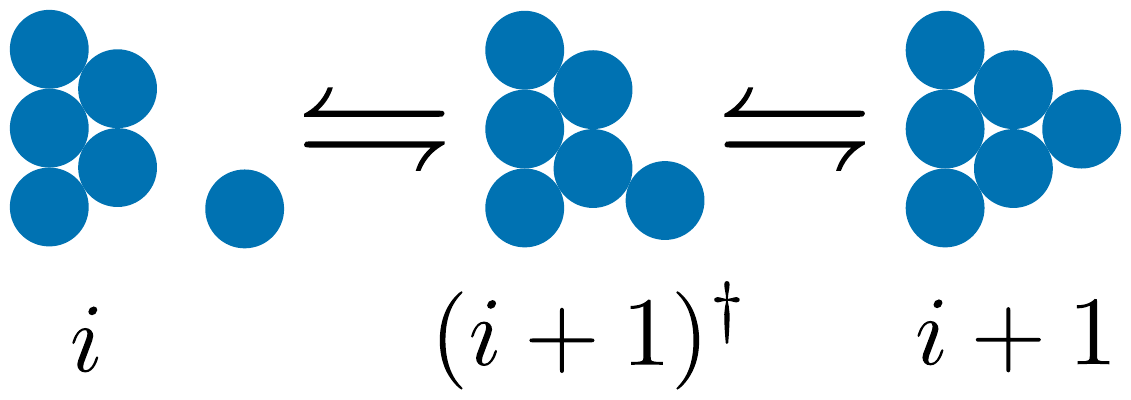}\end{center}
This two-step process can be described by the reaction scheme
\begin{equation}
  i \xrightleftharpoons[v_\infty/dK]{v_\infty/d} (i+1)^\dagger \xrightarrow{\kappa} (i+1).
\end{equation}
The first step involves the attachment of a new colloid by fewer than $z/2$ bonds at the diffusion-limited rate $v_\infty / d$, which we express in terms of the growth velocity in the direction normal to the surface of the cluster, $v_\infty$.
The diffusion-limited rate follows the scaling relation~\cite{palberg1999crystallization}
\begin{equation}
  \label{eq:vinf}
  \frac{v_\infty}{d} \sim \frac{D}{\lambda^2} \simeq D\rho^{2/3},
\end{equation}
where $D$ is the self-diffusion coefficient and $\lambda \sim \rho^{-1/3}$ is the mean free path length in the colloidal gas phase.
Dissociation occurs according to detailed balance and the equilibrium constant $K$.
The second step, associated with the rate $\kappa$, reflects the need for a colloid to roll into place once in the state $(i+1)^\dagger$; rolling can be described by diffusion in two dimensions~\cite{Wang2015,Miranda2018SoftMatter}, and $1/\kappa$ is the characteristic time required for a colloid to roll from one lattice position to another.
At coexistence between the colloidal gas and crystal phases, the effective rate (i.e., the reciprocal of the mean first passage time) for transitioning from state $i$ to state $i+1$ is
\begin{equation}
  \label{eq:katt}
  k_{\mathrm{att}} = \frac{\kappa}{(1 + K^{-1}) + \kappa d/v_\infty}.
\end{equation}
Importantly, $k_{\mathrm{att}}$ does not depend strongly on the diffusion-limited rate when $K \ll 1$ and $\kappa \lesssim v_\infty / d$, i.e., when the activated intermediate state is unstable and the rolling rate is comparable to or slower than the diffusion-limited attachment rate.

Again invoking the cell model~\cite{charbonneau2007gas}, we estimate the equilibrium constant for attaching a colloid at a non-crystalline position according to
\begin{equation}
  \label{eq:KS}
  K(S) \approx S (\Omega / \delta^3) \exp(\Delta \epsilon^\dagger / k_{\mathrm{B}}T),
\end{equation}
where $\Omega \approx \delta d^2$ is the free volume in the intermediate state $(i+1)^\dagger$, $\Delta \epsilon^\dagger \simeq (z/2 - z') \epsilon$, and $z'$ is the number of bonds formed in the intermediate state.
Taking $\epsilon \approx -4.5 k_{\mathrm{B}}T$ and $z' = 1$, we estimate $K(S = 1) \approx 10^{-3}$ for the attachment of a particle to a small nucleus with high surface curvature, since in this case the adsorbed particle is likely to make only a single bond with the cluster.
When attaching to a nearly flat crystalline interface, an adsorbed particle may make a greater number of bonds, leading to a larger value for $K$.

\subsection{Classical nucleation theory}

We assume that the nucleation rate density for homogeneous nucleation follows the classical nucleation theory (CNT) form~\cite{sear2007nucleation}
\begin{equation}
  k_{\mathrm{n}} = k_{\mathrm{n},0}(\rho,T) \exp\left(-\Delta G^\ddagger/\kT\right),
\end{equation}
where $k_{\mathrm{n},0}$ is referred to as the nucleation-rate prefactor and $\Delta G^\ddagger$ is the free-energy difference between the top of the nucleation barrier and the metastable phase.

\subsubsection{Nucleation barrier}

We assume that the nucleation barrier can be described using the capillarity and spherical-nucleus approximations~\cite{oxtoby1992homogeneous},
\begin{equation}
  \label{eq:DeltaG-nucleation}
  \frac{\Delta G^\ddagger}{\kT} = -\frac{16\pi}{3} \frac{\left[\gamma(T)\rho_{\mathrm{c}}^{-2/3}/\kT\right]^3}{(\log S)^2},
\end{equation}
where $\gamma(T)$ is the temperature-dependent crystal--vapor surface tension.
The density in the crystal phase can be determined from the packing fraction, $\phi$, and the colloid diameter, $d$, such that ${\rho_{\mathrm{c}} = (6 \phi/\pi) d^{-3}}$.
We comment further on the interpretation of the surface tension in \secref{sec:surface-tension}.

\subsubsection{Nucleation rate prefactor}

Following Reference \cite{oxtoby1992homogeneous}, the nucleation prefactor can be written as
\begin{equation}
  k_{\mathrm{n},0} = 2\rho \left[\frac{k_{+,\mathrm{eq}}(i^*)}{A(i^*) \rho_{\mathrm{c}}^{2/3}}\right] \left(\frac{\rho_{\mathrm{c}}^{-2/3}\gamma}{\kT}\right)^{1/2},
\end{equation}
where $i^*$ is the critical cluster size, $A(i^*)$ is the critical cluster surface area, and $k_{+,\mathrm{eq}}$ is the rate at which the cluster grows by one particle at coexistence between the crystal and gas phases.
We ignore the distinction between the two colloidal species here for simplicity.
We can replace the term in square brackets with $k_{\mathrm{att}}$ evaluated at coexistence, leading to
\begin{equation}
  \label{eq:knuc0}
  \frac{k_{\mathrm{n},0}}{\rho} = 2 \times k_{\mathrm{att}}(S=1) \times \left(\frac{\rho_{\mathrm{c}}^{-2/3}\gamma}{\kT}\right)^{1/2}.
\end{equation}
As described above, $k_{\mathrm{att}}(S=1)$ is expected to be essentially independent of the concentration, since $K(S=1) \ll 1$ and $\kappa \simeq D\rho_0^{2/3}\!$.
As a result, the nucleation rate density is expected to be proportional to the colloid concentration.
Empirically, we find that such a linear concentration scaling is consistent with the experimental data (see \secref{sec:MLE-singlerate}).
Note that this prediction differs from applications of CNT that do not account for the attachment kinetics and typically predict a stronger density dependence of the nucleation rate prefactor.
For example, the diffusion-limited nucleation rate prefactor for repulsive colloids crystallizing from a dense fluid is expected to scale as $\rho^{5/3}$~\cite{palberg1999crystallization}.

\subsection{Crystal growth}

To describe crystal growth, we follow the presentation in Reference \cite{libbrecht2003growth}.
The diffusion equation at steady state reduces to Laplace's equation,
\begin{equation}
  \label{eq:laplace}
  \frac{\partial \sigma}{\partial t} = 0 = -\nabla^2 \sigma,
\end{equation}
where we've defined
\begin{equation}
  \sigma \equiv \frac{\rho - \rho_{\mathrm{eq}}}{\rho_{\mathrm{eq}}} = S - 1.
\end{equation}
The growth velocity, $v$, of a roughly spherical crystal of radius $R$ is
\begin{equation}
  v = \frac{D}{\rho_{\mathrm{c}}} \frac{\partial \rho(R)}{\partial r} = \frac{D \rho_{\mathrm{eq}}}{\rho_{\mathrm{c}}} \frac{\partial \sigma(R)}{\partial r}
\end{equation}
by continuity.
This velocity is assumed to follow a Wilson--Frenkel law~\cite{saito1996statistical},
\begin{equation}
  v = \frac{\alpha v_\infty(R)}{S} \left[\sigma(R) - \frac{2\xi}{R}\right],
\end{equation}
where $\alpha \le 1$ is the attachment coefficient, $v_\infty(R)$ is the diffusion-limited growth velocity at the crystal--vapor interface, and the term involving $\xi \equiv \gamma / \rho_{\mathrm{c}}k_{\mathrm{B}}T$ is the Gibbs--Thomson correction.
Using \eqref{eq:katt}, we can write the supersaturation-dependent attachment coefficient as
\begin{equation}
  \label{eq:alpha-S}
  \alpha(S) = \frac{k_{\mathrm{att}}}{v_\infty / d} = \frac{\kappa d/v_\infty}{\left[1 + K(S)^{-1}\right] + \kappa d / v_\infty}.
\end{equation}
Finally, to account for particle conservation within the droplet, we impose the condition
\begin{equation}
  V_{\mathrm{drop}}\rho_0 - N = 4\pi \int_{V_{\mathrm{drop}}} \! r^2 \rho(r) dr = 4\pi \int_{V_{\mathrm{drop}}} \! r^2 \rho_{\mathrm{eq}} \left[ \sigma(r) + 1 \right] dr,
\end{equation}
where $N$ is the number of colloids in the crystal phase.

If the droplet is large compared to equilibrium crystal size, then we can approximate the solution to this model to yield the growth velocity
\begin{equation}
  \label{eq:velocity}
  v \simeq \frac{\alpha\alpha_{\mathrm{diff}}}{\alpha + \alpha_{\mathrm{diff}}} \left(\frac{v_\infty}{S}\right) \left[ \sigma_0 - \frac{2\xi}{R} - \frac{4\pi R^3}{3V_{\mathrm{drop}}} \left(\frac{\rho_{\mathrm{c}}}{\rho_{\mathrm{eq}}}\right) \right],
\end{equation}
where $\sigma_0 = S_0 - 1$ describes the initial colloidal gas concentration, and the dimensionless parameter $\alpha_{\mathrm{diff}}$ is
\begin{equation}
  \alpha_{\mathrm{diff}} \equiv \frac{\rho_{\mathrm{eq}} D S}{\rho_{\mathrm{c}} v_\infty R}.
\end{equation}
Using the scaling relation for $v_\infty$ given in \eqref{eq:vinf}, we can write $\alpha_{\mathrm{diff}} = 1 / \rho_{\mathrm{c}}d \lambda R = (\pi/6\phi) d^2 / \lambda R$.
Note that \eqref{eq:velocity} is valid for an equimolar mixture of two colloidal species, where $\sigma_0$ is the same for each colloidal species.

\subsubsection{Post-critical nucleus growth}
\label{sec:postcriticalGrowth}

If $\alpha_{\mathrm{diff}} \gg \alpha$, then the growth velocity reduces to
\begin{equation}
  \label{eq:lagtime-velocity-WF}
  v \simeq \alpha v_{\infty,0} \left[1 - \left(1 + \frac{2\xi}{R}\right)S_0^{-1}\right],
\end{equation}
where $v_{\infty,0}/d$ is the diffusion-limited growth rate at the initial colloid concentration.
This condition is the situation immediately following a nucleation event.

Once nucleation has occurred, the nucleus must grow to a certain size before it can be detected in the experiment.
We assume that this smallest detectable cluster contains $N_{\mathrm{min}} \approx$~50--200 particles (see \secref{sec:imageDataAnalysis}).
We then expect that the first passage time for the post-critical nucleus to grow to $N_{\mathrm{min}}$ particles will be inversely proportional to
\begin{equation}
  \label{eq:lagtime-velocity}
  v / d \simeq \left[2 + K(S_0)^{-1}\right]^{-1} \times (v_{\infty,0}/d) \left(1 - 2S_0^{-1}\right),
\end{equation}
where we have substituted the critical nucleus size into the Wilson--Frenkel law and approximated ${\kappa d / v_{\infty,0} \approx 1}$.
Note that these two approximations are not strictly necessary, but they reduce the number of fitting parameters needed below.

\begin{figure}[t]
  \begin{center}
    \includegraphics[width=\textwidth]{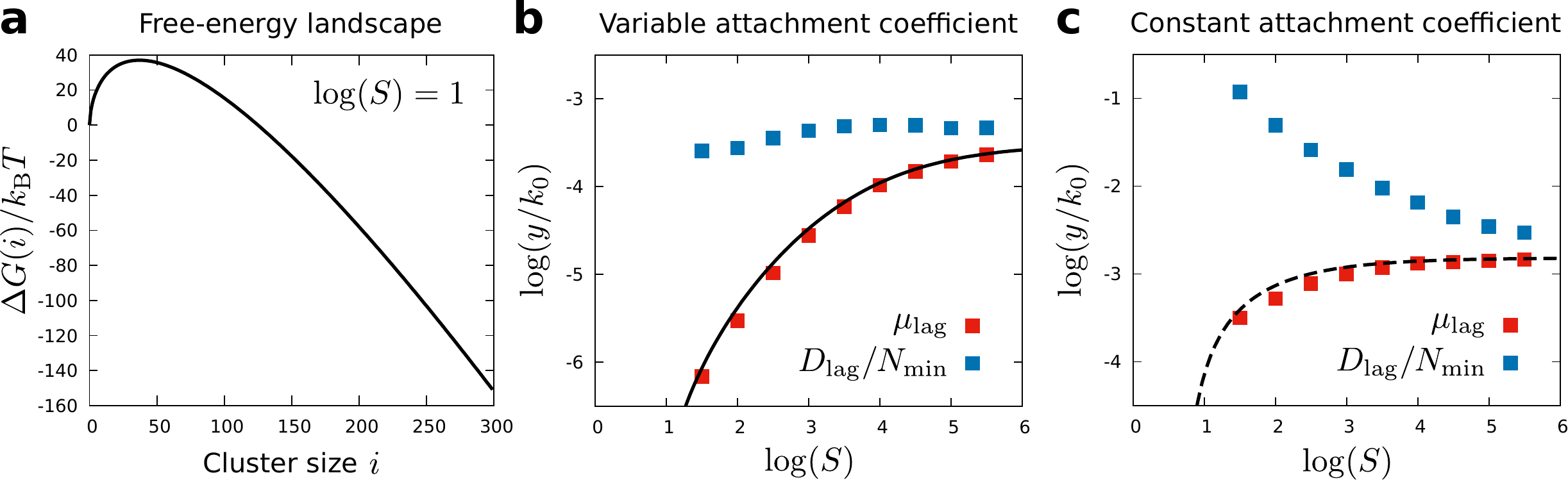}
    \caption{Simulations of post-critical nucleus growth.  (\textbf{a}) The model free-energy landscape, \eqref{eq:modelLandscape}, for parameter values $\theta = 10$ and $\log S = 1$.  (\textbf{b}) Fits of the mean first passage time distributions to the Inverse Gaussian (IG) distribution with the supersaturation-dependent attachment coefficient $\alpha$ given by \eqref{eq:alpha-S}.  Both the drift and diffusion parameters, $\mu_{\text{lag}}$ and $D_{\text{lag}}$, are shown relative to the fundamental rate in the kMC simulations, $k_0$.  The solid black line is the prediction given by \eqref{eq:lagtime-velocity}.  (\textbf{c}) Fits of the mean first passage time distributions to the IG distribution with a supersaturation-independent $\alpha$. The dashed black line is the prediction given by the Wilson--Frenkel law with constant $\alpha$. \label{fig:kmc-growth}}
  \end{center}
\end{figure}

To examine the consequences of a supersaturation-dependent attachment coefficient on the post-critical dynamics, we perform kinetic Monte Carlo (kMC) simulations of the growth process of a post-critical nucleus on a model free-energy landscape,
\begin{equation}
  \label{eq:modelLandscape}
  \frac{\Delta G}{k_{\mathrm{B}}T} = \theta i^{2/3} - (\log S) i,
\end{equation}
where $\theta \equiv (36\pi)^{1/3} \rho_{\mathrm{c}}^{-2/3} \gamma / k_{\text{B}} T$.
The simulations generate random walks on this landscape with forward rates (i.e., $i \rightarrow i+1$) given by $k_+(i) = \alpha i^{2/3} \simeq [2 + S^{-1}K(1)^{-1}]^{-1} i^{2/3}$ and the reverse rates determined via detailed balance.
\figref{fig:kmc-growth} presents simulation results using a dimensionless surface tension $\theta = 10$ and an equilibrium constant at coexistence of $K(1) = 0.02$.
Using 10,000 kMC trajectories, we measure the time required for a critical nucleus to grow to $N_{\mathrm{min}} = 300$ particles without returning to a cluster size of 1 (\figref{fig:kmc-growth}a).
The resulting ``lag time'' distributions are reasonably well described by the Inverse Gaussian (IG) distribution,
\begin{equation}
  \IG(t; \mu_{\mathrm{lag}}, D_{\mathrm{lag}}) = \frac{1}{\sqrt{4\pi (D_{\mathrm{lag}}^2) t^3}} \exp\left\{-\frac{[(\mu_{\mathrm{lag}}) t - 1]^2}{4D_{\mathrm{lag}}t}\right\},
\end{equation}
where $\mu_{\mathrm{lag}}$ and $D_{\mathrm{lag}}$ are the drift and diffusion coefficients for this first-passage process; the lag time discussed in the main text is $\tau_{\mathrm{lag}}\equiv1/\mu_{\mathrm{lag}}$.

Fits to the IG distribution verify that the drift coefficient is well described by \eqref{eq:lagtime-velocity} when a supersaturation-dependent $\alpha$ is used in the expression for $k_+$ in the kMC simulations (\figref{fig:kmc-growth}b).
Furthermore, the best-fit diffusion coefficient is roughly constant, with a value approximately equal to $N_{\text{min}}$ times the asymptotic value of the best-fit drift coefficient; thus, we can extract $N_{\text{min}}$ from the IG fits by calculating ${N_{\text{min}} \approx D_{\mathrm{lag}} / \mu_{\mathrm{lag}}}$ for $\log(S) \gtrsim 5$.
All of these features are consistent with the lag time fits to the nucleation data, as discussed below in \secref{sec:MLE-singlerate}.
By contrast, simulations in which $\alpha$ is treated as a supersaturation-independent constant, resulting in a drift coefficient proportional to the Wilson--Frenkel law given by \eqref{eq:lagtime-velocity-WF}, qualitatively disagree with the lag time fits to the nucleation data.
In this case, the inferred drift coefficient rapidly approaches its asymptotic value for $\log(S) \gtrsim 2$, and the inferred diffusion coefficient varies strongly with the supersaturation (\figref{fig:kmc-growth}c).

\subsubsection{Diffusion-limited crystal growth}

If $\alpha_{\mathrm{diff}} \lesssim \alpha$, then the growth velocity is approximately
\begin{equation}
  v \simeq \frac{D}{\rho_{\mathrm{c}} R} \left(1 - \frac{\alpha_{\mathrm{diff}}}{\alpha}\right) \left(\rho_0 - \rho_{\mathrm{eq}} - \frac{4\pi R^3\rho_{\mathrm{c}}}{3V_{\mathrm{drop}}}\right).
\end{equation}
Using the relation $N = 4\pi R^3 \rho_{\mathrm{c}} / 3$, we can write the diffusion-limited growth law as
\begin{equation}
  \label{eq:diffusion-limited}
  \frac{dN}{dt} \simeq 4\pi R D \left(1 - \frac{\alpha_{\mathrm{diff}}}{\alpha}\right)\left(\rho_0 - \rho_{\mathrm{eq}} - \frac{N}{V_{\mathrm{drop}}}\right).
\end{equation}
There are two reasons why we expect the growth process to eventually enter a regime in which $\alpha_{\mathrm{diff}} < \alpha$.
First, $\alpha_{\mathrm{diff}}$ decreases as $1/R$ as the crystal grows.
Second, the equilibrium constant $K(S)$ can be expected to increase as the radius of curvature of the crystal--vapor interface increases, leading to an increase in the typical number of bonds formed by an attached but as yet uncrystallized particle, $z'$.
Thus, this model predicts a switch from kinetics-limited growth in the immediate aftermath of a nucleation event to diffusion-limited growth at later times.

Integrating \eqref{eq:diffusion-limited} when $N/V_{\mathrm{drop}} \ll \rho_0 - \rho_{\mathrm{eq}}$, we find that the number of crystallized particles should scale as $N \sim t^{3/2}$.
Once $N/V_{\mathrm{drop}}$ becomes comparable to $\rho_0 - \rho_{\mathrm{eq}}$, the difference between the colloidal gas density and the gas density at equilibrium, $(\rho_0 - N/V_{\mathrm{drop}}) - \rho_{\mathrm{eq}}$, is expected to decay exponentially.
These predictions are confirmed by fitting \eqref{eq:diffusion-limited} to the growth data (see Figure~1F and Figure~3D in the main text).

\section{Fitting the experimental data}

\subsection{Maximum likelihood estimation of nucleation rates and lag times}

We use maximum likelihood estimation (MLE) to analyze the nucleation and post-critical growth behavior.
Given a list of times, ${\{t_j\}}$, at which a supercritical nucleus was first observed in a droplet, we seek to maximize the log-likelihood function
\begin{equation}
  \log L = \sum_{j = 1}^{n_{\mathrm{obs}}} \log p(t_j) + (n_{\mathrm{tot}} - n_{\mathrm{obs}}) \log p(t > t_{\mathrm{max}}),
\end{equation}
where $p(t)$ is the model probability of observing a supercritical nucleus at time $t$, ${n_{\mathrm{obs}}}$ is the number of observations (${j = 1,\ldots,n_{\mathrm{obs}}}$), and ${n_{\mathrm{tot}}}$ is the total number of droplets in the experiment.

\subsubsection{Single nucleation rate}
\label{sec:MLE-singlerate}

To begin, we assume that nucleation in all droplets can be described by a single rate, $k_{\mathrm{n}}$.
In this case, following the discussion in \secref{sec:postcriticalGrowth}, the probability of observing a supercritical nucleus in an experiment at time $t$ is
\begin{equation}
  p(t) = \int_0^t d(\Delta t) \IG(\Delta t) k_{\mathrm{n}} e^{-k_{\mathrm{n}}(t - \Delta t)}.
\end{equation}
The probability that a supercritical nucleus is not observed in a droplet before the end of the experiment at time $t_{\mathrm{max}}$ is
\begin{align}
  p(t > t_{\mathrm{max}}) &= \int_{t_{\mathrm{max}}}^\infty dt \int_0^t d(\Delta t) \IG(\Delta t; \mu_{\mathrm{lag}}, D_{\mathrm{lag}}) k_{\mathrm{n}} e^{-k_{\mathrm{n}}(t-\Delta t)} \\
  &\simeq \int_{t_{\mathrm{max}}}^\infty dt\, k_{\mathrm{n}} e^{-k_{\mathrm{n}}(t - t^*)} = e^{-k_{\mathrm{n}}(t_{\mathrm{max}} - t^*)},
\end{align}
where we have defined
\begin{equation}
  t^* \equiv \frac{1}{k_{\mathrm{n}}} \log \int_0^\infty d(\Delta t) e^{k_{\mathrm{n}}\Delta t} \IG(\Delta t; \mu_{\mathrm{lag}}, D_{\mathrm{lag}}).
\end{equation}

\begin{figure}
  \begin{center}
    \includegraphics[width=0.75\textwidth]{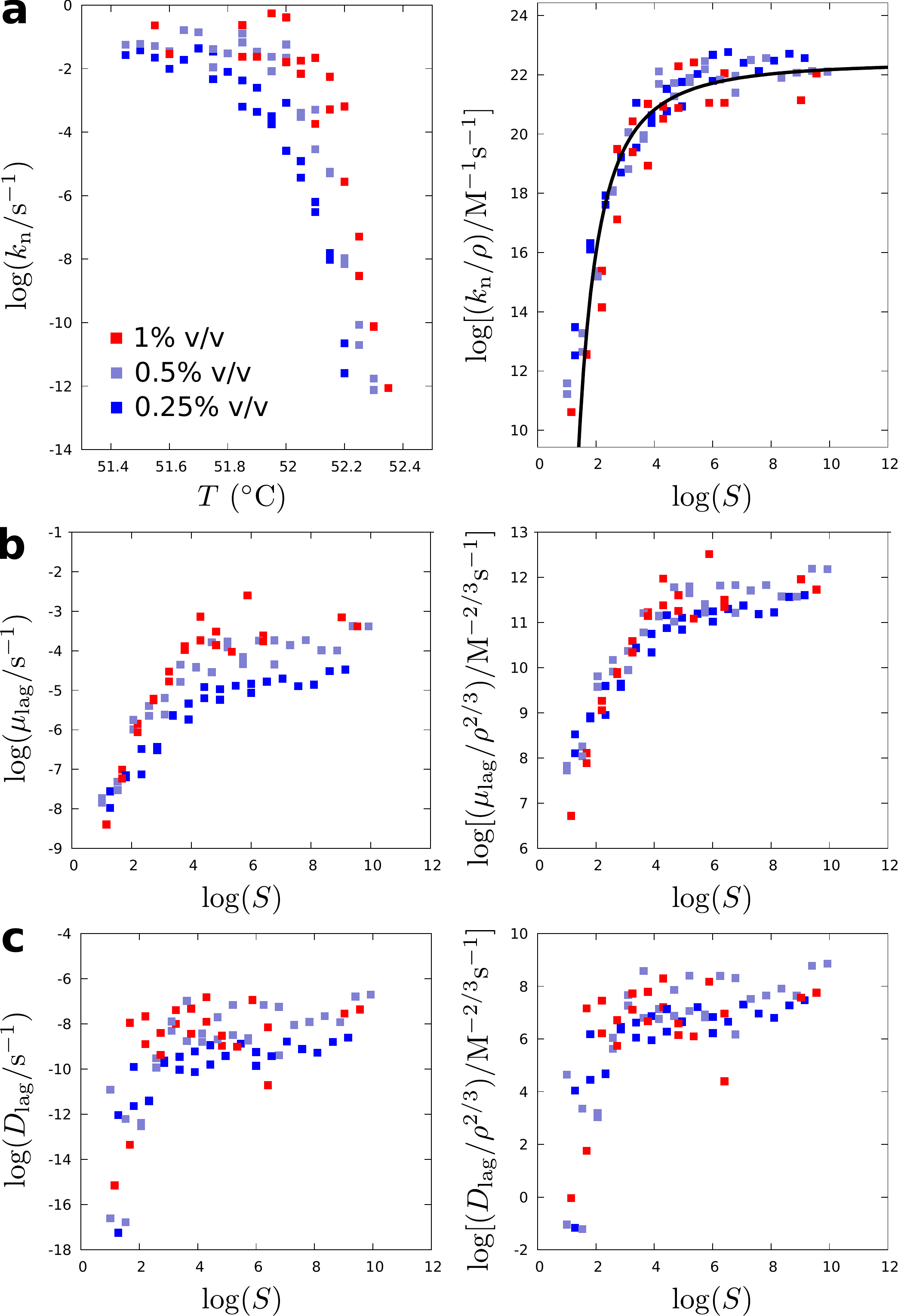}
  \end{center}
  \caption{Nucleation and early-time growth parameters determined from fits to the single-nucleation-rate MLE model described in \secref{sec:MLE-singlerate}.  The fitting parameters $k_{\mathrm{n}}$, $\mu_{\mathrm{lag}}$, and $D_{\mathrm{lag}}$ are determined separately for each quench.  Panels on the left show the raw data for the (\textbf{a}) nucleation rate, (\textbf{b}) lag time drift coefficient, and (\textbf{c}) lag time diffusion coefficient, while panels on the right show the rescaled data.  In panel \textbf{a}(\textit{right}), the black curve shows a fit to \eqref{eq:knB} with $k_{\mathrm{n,0}} = \exp(22.4)$ and $B = 25.8$.  The fit to the rescaled data in panel \textbf{b}(\textit{right}) is shown in \figref{fig:MLE-varconc-growth}a.  The plateau in panel \textbf{c}(\textit{right}) is consistent with the prediction of \figref{fig:kmc-growth} and is thus used to fix a supersaturation-independent $D_{\text{lag}}$ in the variable-concentration fits described in \secref{sec:MLE-variablerate}. \label{fig:MLE-singlerate}}
\end{figure}

We fit this MLE model to the data from each concentration and temperature with $k_{\mathrm{n}}$, $\mu_{\mathrm{lag}}$, and $D_{\mathrm{lag}}$ as fitting parameters.
\figref{fig:MLE-singlerate} shows the raw parameters that we obtain, along with the predicted concentration scalings.
The data collapse for $k_{\mathrm{n}}$ confirms that the nucleation rate prefactor is roughly linear in concentration and that the inferred nucleation barrier heights are well described by CNT.
From the asymptotic behavior at high supersaturation, we find that $k_{\mathrm{n},0}/V_{\mathrm{drop}}\rho \simeq V_{\mathrm{drop}}^{-1} \times \exp(22.4)\,\mathrm{M}^{-1}\mathrm{s}^{-1} \simeq 1 \times 10^{-4}\,\mathrm{s}^{-1}$.
Given that $\gamma \rho_{\mathrm{c}}^{-2/3}$ is in the range of $1$ to $2k_{\mathrm{B}}T$, we use \eqref{eq:knuc0} to estimate that $k_{\mathrm{att}}(S=1) \simeq 1 \times 10^{-4}\,\mathrm{s}^{-1}$, which is close to our estimate from \eqref{eq:KS}.
Thus, the inferred nucleation-rate prefactor is consistent with our attachment-kinetics model to within an order of magnitude.

Analysis of the lag times shows that both $\mu_{\mathrm{lag}}$ and $D_{\mathrm{lag}}$ collapse onto master curves when scaled by $\rho^{2/3}$, as expected from \eqref{eq:lagtime-velocity}.
Furthermore, the inferred diffusion coefficients are roughly constant as a function of supersaturation, which is consistent with the predictions of the kMC simulations shown in \figref{fig:kmc-growth}.
We also confirm that $D_{\mathrm{lag}} / \mu_{\mathrm{lag}}$ in the asymptotic regime of high supersaturation is of order 100, which is consistent with a minimum observable cluster size, $N_{\mathrm{min}}$, of approximately 50--200 colloidal particles.
Note that the uncertainty in the diffusion coefficients is large at low supersaturation because crystallization events were not observed in most of these droplets within the duration of the experiment.

\subsubsection{Accounting for variable droplet concentrations}
\label{sec:MLE-variablerate}

\begin{table}
  \begin{center}
    \begin{tabular}{c|c|c|c}
      \textbf{Nominal concentration (\% v/v)} & \textbf{Mean concentration (M)} & \textbf{Standard deviation (M)} & \textbf{$\Delta$} \\
      \hline \hline
      0.25 & $3.33\times10^{-11}$ & $0.30\times10^{-11}$ & $0.090$ \\
      0.50 & $7.24\times10^{-11}$ & $0.49\times10^{-11}$ & $0.067$ \\
      1.00 & $14.19\times10^{-11}$ & $0.58\times10^{-11}$ & $0.041$
    \end{tabular}
  \end{center}
  \caption{Concentration measurements for the three nominal concentrations (cf.~\figref{fig:scattering}). \label{tab:concentrations}}
\end{table}

Now we account for variations in the colloid concentrations from droplet to droplet (see \tabref{tab:concentrations}).
We assume that the concentration in a randomly selected droplet follows a Gaussian distribution with mean $\bar\rho$ and standard deviation $\Delta\rho$.
We define the mean supersaturation to be $\bar S \equiv \bar \rho / \rho_{\mathrm{eq}}(T)$ and the relative variation in concentrations to be $\Delta \equiv \Delta\rho / \bar \rho$.
Assuming a CNT rate of the form
\begin{equation}
  \label{eq:knB}
  k_{\mathrm{n}}(\rho, S, B) = k_{\mathrm{n},0} \rho \exp\left[\frac{-B}{(\log S)^2}\right],
\end{equation}
where $B$ is a fitting parameter, the probability that a nucleation event occurs at time $t$ in any of the droplets is
\begin{equation}
  p_{\mathrm{nuc}}(t; \bar S, \Delta, B) = \int_{1/\bar S}^\infty d\zeta \left(\frac{e^{-(\zeta-1)^2/2\Delta^2}}{\sqrt{2\pi\Delta^2}}\right) k_{\mathrm{n}}(\zeta\rho_{\mathrm{eq}}\bar S, \zeta\bar S, B) e^{-k_{\mathrm{n}}(\zeta\rho_{\mathrm{eq}}\bar S, \zeta\bar S, B) t},
\end{equation}
where $\zeta \equiv \rho / \bar \rho$.
The lower integration limit reflects the fact that nucleation can only occur when ${\rho > \rho_{\mathrm{eq}}(T)}$.
The probability of observing a supercritical nucleus at time $t$ is
\begin{equation}
  p(t) = \int_0^t d(\Delta t)\IG(\Delta t)p_{\mathrm{nuc}}(t - \Delta t),
\end{equation}
while the probability of not observing nucleation prior to the end of the experiment is
\begin{equation}
  p(t > t_{\mathrm{max}}) = p_{\mathrm{never}} + \int_{t_{\mathrm{max}}}^\infty dt\, p(t),
\end{equation}
where $p_{\mathrm{never}}$ is the probability that the concentration in a drop is at or below the coexistence concentration, ${\rho \le \rho_{\mathrm{eq}}(T)}$,
\begin{equation}
 p_\textrm{never} =  \int_{-\infty}^{1/\bar S} d\zeta \left(\frac{e^{-(\zeta-1)^2/2\Delta^2}}{\sqrt{2\pi\Delta^2}}\right).
\end{equation}
In practice, we find that $p_{\mathrm{never}}$ is negligible for the values of $\bar S$ and $\Delta$ considered.

\begin{figure}
  \begin{center}
    \includegraphics[width=0.75\textwidth]{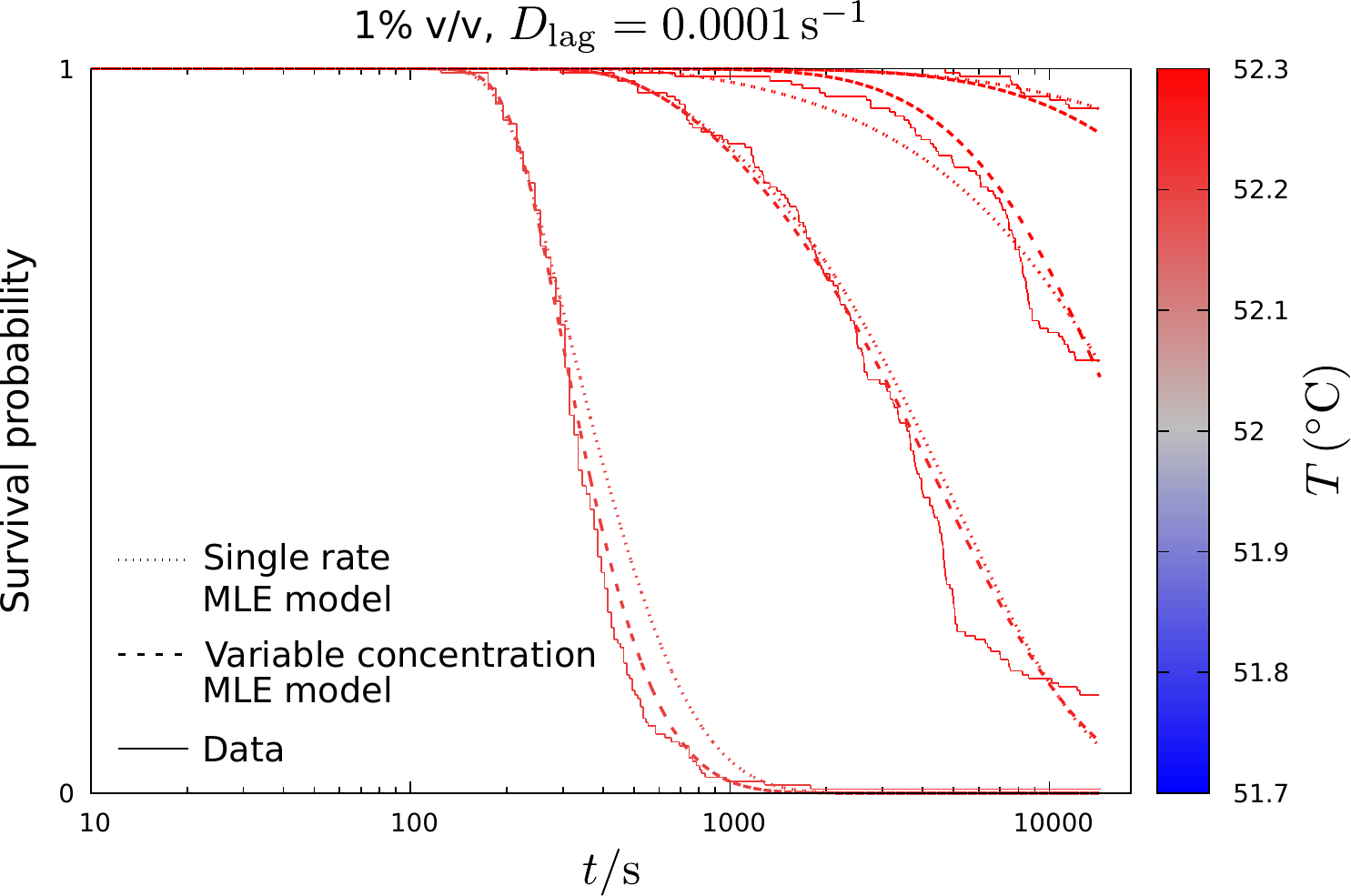}
  \end{center}
  \caption{Comparison of the single-nucleation-rate and variable-concentration MLE models at low supersaturation.  In this regime, roughly $\log S \lesssim 2$, variations in the supersaturation across droplets at the same nominal concentration are not insignificant, and the differences between the best fits of the two MLE models are apparent. \label{fig:compare-MLE-models}}
\end{figure}

We fit this variable-concentration MLE model to the experimental data at all concentrations and temperatures.
To reduce the number of fitting parameters, we impose a concentration-dependent $D_{\mathrm{lag}}$ and a single constant value of $k_{\mathrm{n},0}/\rho$, as determined from \figref{fig:MLE-singlerate}c.
The remaining fitting parameters are thus $B$ and $\mu_{\mathrm{lag}}$.
The results of this fitting procedure are shown in Figures~2 and 3 in the main text.
We note that the difference between the two MLE models is most apparent at low average supersaturation.
In this regime, the variation among CNT rates in droplets under the same nominal conditions is relatively large, and thus deviations from single-exponential nucleation rates become apparent.
For visual comparison of these two models, representative predicted survival functions are shown along with the experimental survival functions in \figref{fig:compare-MLE-models}.

\begin{figure}
  \begin{center}
    \includegraphics[width=0.75\textwidth]{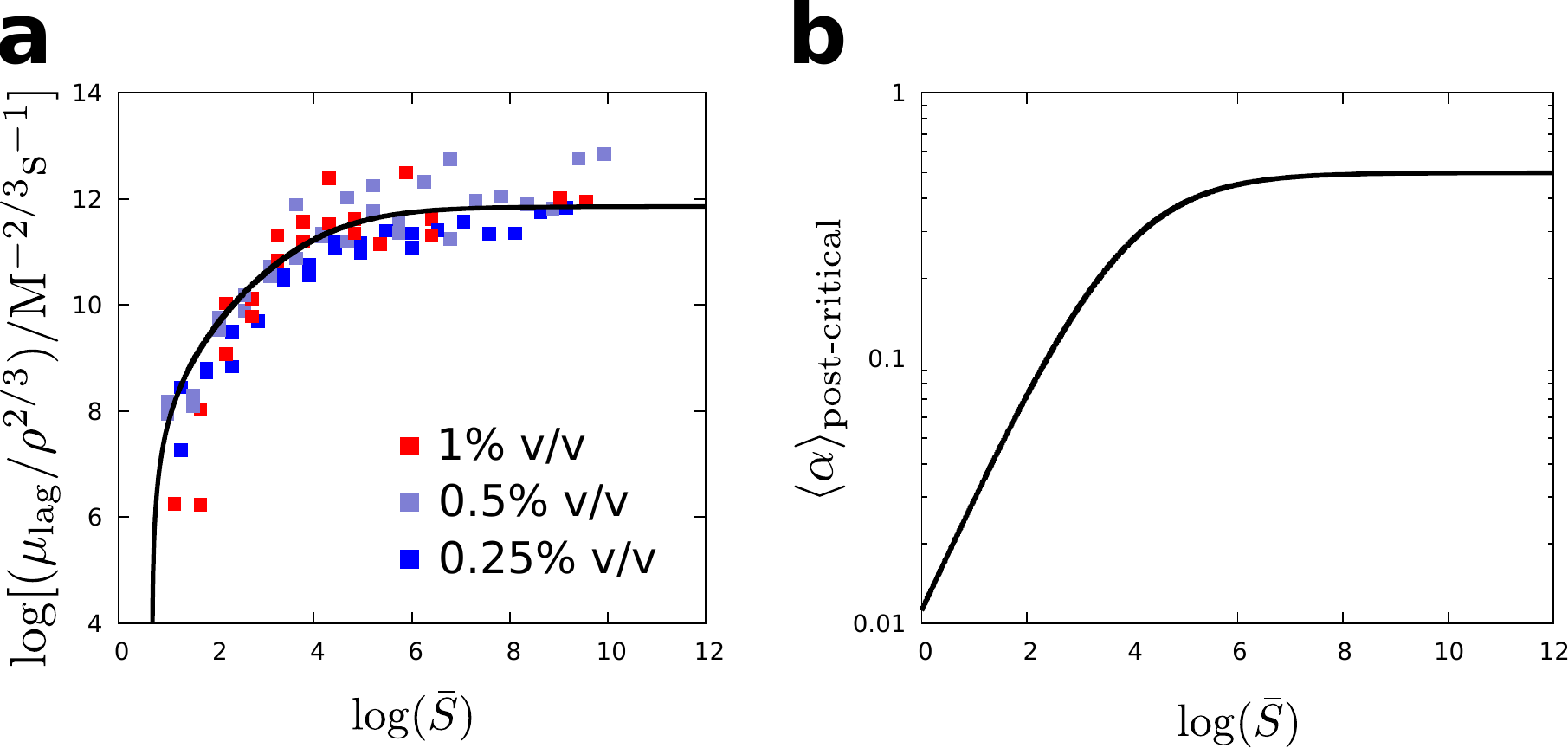}
    \caption{The rescaled lag time drift coefficient obtained from fitting the variable-concentration MLE model described in \secref{sec:MLE-variablerate}.  (\textbf{a}) The data in this panel correspond to the lag times shown in Figure~3c in the main text.  The black line is the fit to \eqref{eq:lagtime-velocity}, which yields the estimate $K/S = 0.01$.  (\textbf{b}) Given this estimate for $K/S$, we plot the inferred supersaturation-dependent attachment coefficient, which represents an average of the attachment coefficient over post-critical nuclei with fewer than $N_{\text{min}}$ particles.  \label{fig:MLE-varconc-growth}}
  \end{center}
\end{figure}

We find that fixing a supersaturation-independent value of $D_{\mathrm{lag}}$, as described above, tends to reduce the noise in the estimation of the mean lag times, $\tau_{\text{lag}} = 1 / \mu_{\text{lag}}$.
The lag time drift coefficients shown in \figref{fig:MLE-varconc-growth}a correspond to the mean lag times shown in Figure~3c in the main text.
The asymptotic value that we extract at high supersaturations, approximately $\mu_{\mathrm{lag}} / \rho^{2/3} \simeq 3\times10^{5}\,\mathrm{M}^{-2/3}\mathrm{s}^{-1}$, compares favorably with the prediction of \eqref{eq:lagtime-velocity}, $\mu_{\mathrm{lag}} / \rho^{2/3} \sim \alpha D N_{\mathrm{min}}^{1/3} \simeq 5\times10^{4}\,\mathrm{M}^{-2/3}\mathrm{s}^{-1}$ in the limit $S \rightarrow \infty$.
We are also able to extract the equilibrium constant $K \simeq 0.01 S$ by fitting $\mu_{\mathrm{lag}} / \rho^{2/3}$ to \eqref{eq:lagtime-velocity}.
From this fit, we obtain inferred values of the attachment coefficient for post-critical nucleus growth, which are plotted in \figref{fig:MLE-varconc-growth}b.
The range of values obtained here, $0.01 \le \langle\alpha\rangle_{\text{post-critical}} \le 1$, is consistent with the predicted cross-over from kinetics limited attachment at the critical nucleus to diffusion-limited growth at later times, when $R \gg d$ and thus $\alpha_{\mathrm{diff}} = (\pi/6\phi) d^2 / \lambda R \ll 1$.
We note that although the value of $K/S$ obtained here is two orders of magnitude larger than the value of $k_{\mathrm{att}}(S=1)$ obtained from the nucleation rate prefactor in \secref{sec:MLE-singlerate}, this difference is expected, since the inference from the lag time data reflects an average over cluster sizes from the critical nucleus up to $N_{\mathrm{min}}$.
Thus, the larger average value of $K/S$ obtained from the lag time data is consistent with \eqref{eq:KS}, since the typical value of $z'$ in the intermediate state of the attachment process increases as a post-critical nucleus grows and the curvature of the crystal--vapor interface decreases.

\subsection{Interpretation of inferred surface tension values}
\label{sec:surface-tension}

According to the capillarity approximation, the quantity $[B(T) / (16\pi/3)]^{1/3}$ should be interpreted as the dimensionless surface tension, $\gamma(T) \rho_{\mathrm{c}}^{-2/3}\! / \kT$.
Furthermore, at sufficiently low temperatures, $\gamma(T) \rho_{\mathrm{c}}^{-2/3} \!/ \kT$ should be roughly equal to the free energy per unit surface area associated with half the missing ``bonds'' on a planar crystal--vapor interface.
Thus, we expect
\begin{equation}
  \label{eq:gamma-lowT}
  \frac{\gamma(T) \rho_{\mathrm{c}}^{-2/3}}{\kT} \simeq -\frac{z_{\mathrm{surf}} \epsilon}{2\kT},
\end{equation}
where $z_{\mathrm{surf}}$ is the number of missing bonds within an interfacial region of area $d^2$.
Figure~2f in the main text shows that this quantity is indeed roughly linear with respect to temperature, which is consistent with the temperature dependence predicted by \eqref{eq:epsilon} if ${1 \lesssim z_{\mathrm{surf}} \lesssim 2}$.
Furthermore, it appears that $\gamma$ extrapolates to zero near $53^\circ\text{C}$; at this temperature, the cell model breaks down, since \eqref{eq:rhoeq} predicts that the equilibrium gas density is comparable to $\rho_{\text{c}}$.
This behavior is compatible with the expected phase diagram for hard-sphere square-well crystallization~\cite{charbonneau2007gas}.

Nonetheless, the absolute values of $\gamma(T) \rho_{\mathrm{c}}^{-2/3} / \kT$ inferred in this way are slightly lower than those predicted by \eqref{eq:gamma-lowT}.
One likely culprit is the uncertainty associated with the interaction-range parameter $\delta$ used in predicting the absolute value of the binding free energy, as noted in \secref{sec:thermodynamics}.
Another contribution missing from this argument is the conformational entropy of a small critical nucleus, which may lead to substantial deviations from the low-temperature capillarity approximation.
By applying the nucleation theorem, $i^* = \partial \Delta G^\ddagger / \partial \mu$, to the inferred barrier heights shown in Figure~2e in the main text, we estimate that the critical nucleus contains on the order of 10 colloids when $1 \lesssim \log(S) \lesssim 4$.
Thus, substantial conformational fluctuations are likely to result in a roughly temperature-independent contribution to $\gamma/k_{\text{B}}T$ obtained by fitting CNT, which assumes the capillarity approximation, to the data.

A third possibility is that the surface tension is being underestimated due to variations in the grafting density of DNA strands on the surface of the colloidal particles.
These variations lead to a distribution of binding free energies, meaning that \eqref{eq:epsilon} only captures the mean binding free energy and not the interactions between all pairs of particles in a small cluster.
In what follows, we show that binding-energy heterogeneity can lead to an apparent surface tension for small clusters that is lower than that of a macroscopic planar interface, and we estimate the magnitude of this effect.
We assume that the bulk free energy per colloid in CNT follows a Gaussian distribution, $p(\Delta\mu)$, with mean $\overline{\Delta\mu}$ and variance $s^2$.
Assuming that a randomly composed cluster of $i$ colloids can be characterized by a classical free energy of the form ${\Delta G/\kT = \theta i^{2/3} - \Delta\mu i}$ and that there are no correlations between $\Delta\mu$ for the various colloids within a cluster, the resulting projected free energy landscape as a function of cluster size is
\begin{equation}
  \frac{\widetilde{\Delta G}(i)}{\kT} = -\log \int d\!\left(\!\frac{\Delta\mu i}{\kT}\!\right) p\!\left(\!\frac{\Delta\mu i}{kT}\!\right) \exp\left(\!\!-\theta i^{2/3} + \frac{\Delta\mu i}{kT}\right) = \theta i^{2/3} - \left[\!\left(\frac{\overline{\Delta\mu}}{\kT}\right) + \frac{1}{2}\!\left(\frac{s}{\kT}\right)^{\!\!\!2} \right] i.\qquad\;\;
\end{equation}
The final term in this equation implies that variations in the bond energies decrease the critical barrier height, since $s^2 > 0$.
The CNT barrier height is now
\begin{equation}
  \frac{\widetilde{\Delta G}^\ddagger}{\kT} = \frac{-16\pi}{3} \frac{(\gamma \rho_{\mathrm{c}}^{-2/3}/\kT)^3}{\left[\log S + (s/\kT)^2/2\right]^2}.
\end{equation}
Next, we estimate $s^2$ by assuming that the colloid--colloid binding free energies are independent.
If we define the ``patch-to-patch'' polydispersity in the grafting density to be $\Delta_{\mathrm{g}}$, then we obtain $s^2 = z(\Delta_{\mathrm{g}}\bar\epsilon)^2$, where $\bar\epsilon$ is the mean binding free energy.
The effective surface tension appearing in \eqref{eq:DeltaG-nucleation} can therefore be written as
\begin{equation}
  \gamma_{\mathrm{eff}} = \gamma \left[1 + \frac{z}{2\log S}\left(\frac{\Delta_{\mathrm{g}}\bar\epsilon}{\kT}\right)^2\right]^{-2/3}.
\end{equation}
Plugging in the estimated values $\Delta_{\mathrm{g}} \simeq 0.1$ and $\bar\epsilon \simeq 4.5\kT$, we find that $\gamma_{\mathrm{eff}}$ may be $20\%$ smaller than $\gamma$ when $\log S \simeq 2$.
Note that this correction is largest at high temperatures, where the supersaturation is low in the experiments.

\subsection{Fitting diffusion-limited growth}

\begin{figure}[t]
    \centering
    \includegraphics[]{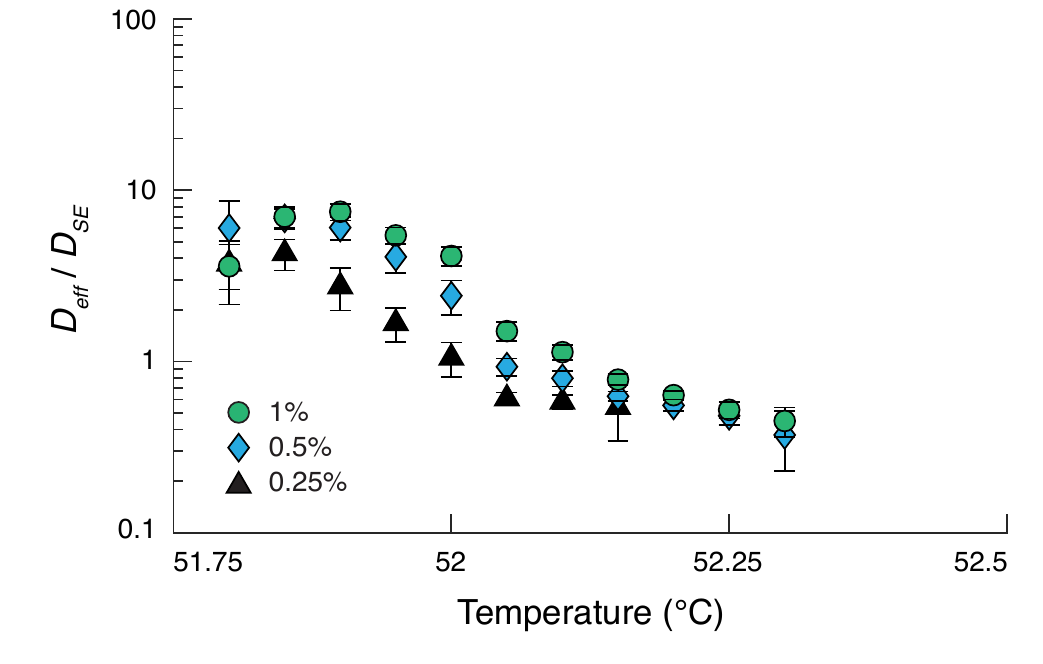}
    \caption{The average effective diffusion coefficient of the crystal growth trajectories relative to the self-diffusion coefficient predicted by Stokes--Einstein, $D_{\mathrm{SE}}$, as a function of temperature. For deep quenches, where many crystals nucleate, $D_{\mathrm{eff}}$ is much larger than the self-diffusion coefficient due to the fact that multiple crystals are growing simultaneously. At higher temperatures, where mostly single crystals form, the effective diffusion coefficient is roughly half of the predicted self-diffusion coefficient.}
    \label{fig:Dcoeff}
\end{figure}

We infer an effective diffusion coefficient for each droplet concentration and temperature by fitting the trajectory of every droplet to the diffusion-limited growth model, \eqref{eq:diffusion-limited}, using an effective diffusion coefficient, $D_{\mathrm{eff}} \equiv D(1 - \alpha_{\mathrm{diff}}/\alpha)$, as a fitting parameter.
For a given concentration and temperature, we take the mean and standard deviation of the fitted diffusion coefficient of all droplets that formed a crystal and plot these results in Figure~\ref{fig:Dcoeff}. For temperatures at which many crystals nucleate, we find that this effective diffusion coefficient is significantly higher than what is predicted by Stokes--Einstein, which we calculate to be $D_{SE}=1.5\times 10^{-12}~\textrm{m}^2/\textrm{s}$.
This observation is likely a consequence of there being multiple crystals growing simultaneously within a single droplet, whereas \eqref{eq:diffusion-limited} assumes a single spherical crystal.
However, for quenches where only a single crystal nucleates in each droplet, we find that the fitted effective diffusion coefficient is within a factor of two of the predicted value, which is consistent with our model if ${\alpha_{\textrm{diff}} / \alpha \lesssim 0.5}$ as expected for diffusion-limited growth. This match between our experiments and our model further supports the conclusion that the late stages of crystal growth are largely diffusion-limited.

\subsection{Predicting distributions of crystals per droplet via simulation}
\label{sec:crystalCount}

Based on our modeling of the nucleation and growth kinetics, we use a simple Monte-Carlo simulation to predict the distribution of the number of crystals grown within each droplet.
To this end, we simulate $1000$ droplets with the same initial nucleation rate $k_{\mathrm{n}}$ determined from the CNT-based analysis described above.
We then evolve the system in discrete time steps of duration $\Delta t = 1\,\mathrm{s}$.
For each time step, we determine whether or not a crystal nucleates within each drop during that time step by approximating the exponential waiting time distribution: A crystal nucleates within droplet $i$ if $\Delta t k_{\mathrm{n}} < p_i$, where $p_i$ is a uniform random number between 0 and 1.
With $\Delta t = 1\,\mathrm{s}$, the probability of nucleation in a given time step for the deepest quenches is around $1\%$ and is significantly lower for shallower quenches.
After the initial nucleation event, the nucleation rate becomes time-dependent, since the growth of the crystal phase depletes the bulk concentration over time.
We account for this effect by numerically integrating the rate of diffusion-limited growth, \eqref{eq:diffusion-limited}, of a spherical crystal over time in a droplet with a radius commensurate with those in the experiment.
We calculate a new nucleation rate for each droplet at every time step after the first nucleation event and perform the same procedure to determine whether another crystal has nucleated.
At the end of the simulation, we obtain the number of crystals that have nucleated within each droplet.

\begin{figure}[H]
    \centering
    \includegraphics[]{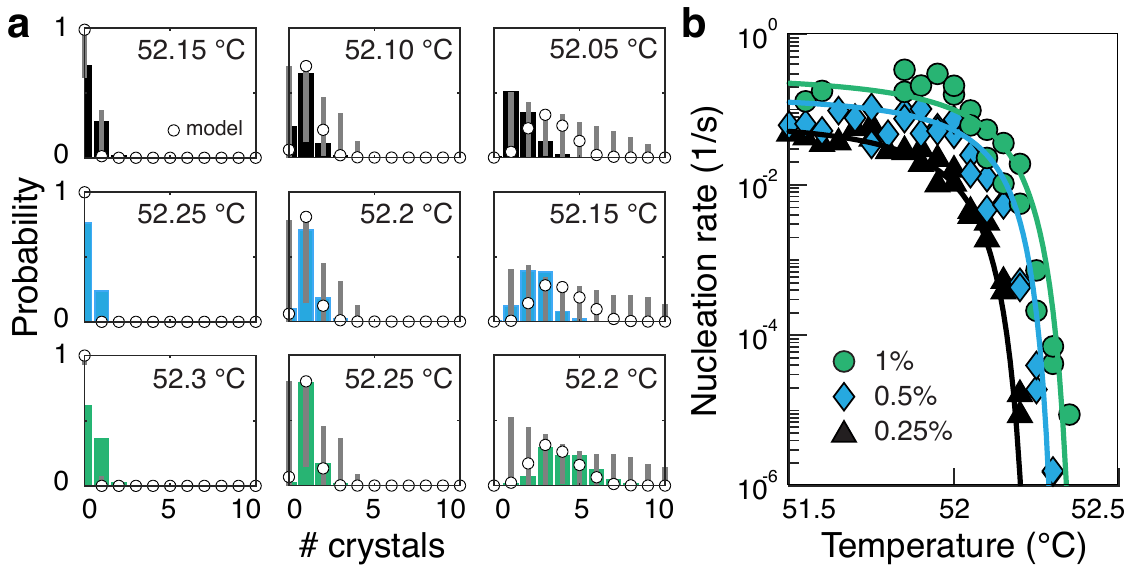}
    \caption{The distributions of the number of crystal domains per droplet for three different colloid concentrations agree favorably with our model predictions of the coupled nucleation and growth of colloidal crystals. (\textbf{a}) shows the probability that a droplet contains a given number of crystals for three different colloids concentrations: 0.25\% (black), 0.5\% (blue), and 1\% (green). For each concentration, we show the distributions for the three highest-temperature quenches where crystals nucleate. Colored bars show experimental data. White circles show model predictions, described in the text, for the nucleation rates given by the solid curves in \textbf{b}. The gray bars show the variability in the probability of forming a given number of crystals due to a variation in the surface tension $\gamma$ of $\pm$10\%. (\textbf{b}) shows the measured nucleation rates (points) and the corresponding model predictions (solid curves).
    }
    \label{fig:crystalCount}
\end{figure}

These simulations confirm the intuition that the crystal morphology depends on a balance of nucleation and growth.
In \figref{fig:crystalCount}, we show the distributions of the number of crystals that form per drop at three different supersaturations, and we note that the fraction of droplets that nucleate multiple crystals increases rapidly with increasing supersaturation.
Overall, the simulations agree well with the experimental results, indicating that CNT is sufficient to explain the morphology of the crystals that self-assemble.

\section{Predicting protocols for growing single crystals}

\subsection{Modeling isothermal crystallization}
We now apply our model of the nucleation and growth dynamics for an individual nucleus to predict the conditions under which we can reliably grow single crystals.
Based on the experiments and modeling discussed above, we know the nucleation rate as a function of the initial concentration and the temperature, $k_{\mathrm{n}}(\rho_0, T)$.
However, once an initial nucleus begins to grow, it will deplete the pool of available monomers and thus alter the nucleation rate elsewhere in the droplet.
To simplify our analysis, we calculate a characteristic time, $\tau_{\mathrm{g}}$, which describes the time required for a single nucleus to grow large enough to suppress nucleation at the initial rate $k_{\mathrm{n}}$ elsewhere in the droplet.
This definition implies that the probability that a new nucleus does \textit{not} appear during the time interval $\tau_{\text{g}}$ is given by $\exp(-k_{\text{n}}\tau_{\text{g}})$.
The probability of growing a single crystal in an isothermal experiment of duration $t_{\mathrm{max}} \gg \tau_{\text{g}}$ is thus given by
\begin{equation}
  \label{eq:isothermalProtocol}
  p_{\mathrm{1x}} = \left\{1 - \exp\left[-k_{\mathrm{n}} t_{\mathrm{max}}\right]\right\} \exp\left[-k_{\mathrm{n}} \tau_{\mathrm{g}}\right].
\end{equation}
This expression states that, to observe a single crystal, the first nucleus must appear within a time interval $t_{\text{max}}$ and no other nucleus can appear within the following characteristic growth time $\tau_{\text{g}}$.

In order to calculate $\tau_{\text{g}}$, we equate the survival probability $\exp(-k_{\text{n}}\tau_{\text{g}})$ to the probability of nucleating a second crystal in the full dynamical growth model, in which the instantaneous nucleation rate in the droplet changes with time.
This condition implies
\begin{equation}
  \exp\left[-k_{\text{n}}(\rho_0,T)\tau_{\text{g}}\right] = \exp\left\{-\int_0^\infty dt\, 4\pi V_{\text{drop}}^{-1} \int_{R(t)}^{R_{\text{drop}}} dr\,r^2k_{\text{n}}[\rho(r,t),T]\right\},
\end{equation}
where $t$ is the time since the appearance of the first nucleus, $R$ is the radius of the first nucleus, $R_{\text{drop}}$ is the droplet radius, and we've assumed that the first nucleus appears at the center of the droplet.
We can therefore express $\tau_{\text{g}}$ as
\begin{equation}
  \label{eq:taug}
  \tau_{\text{g}} =  \int_0^\infty dt \frac{\langle k_{\text{n}} \rangle_t}{k_{\text{n}}(\rho_0,T)},
\end{equation}
where $\langle k_{\text{n}} \rangle_t$ is the average nucleation rate in the droplet at time $t$.
Because $\tau_{\text{g}}$ is much less sensitive to the initial concentration and temperature than the nucleation rate, we treat $\tau_{\text{g}}$ as a constant in the single-crystal probability calculations that follow.

We compare the predictions of this theory with the experimentally observed fraction of single crystals at a nominal 0.5\% v/v concentration in Figure~4a of the main text.
Assuming that crystal growth must have entered the diffusion-limited regime in order to suppress nucleation far away from the first nucleus, we use \eqref{eq:laplace} to calculate $\rho(r,t)$.
In this way, we estimate ${\tau_{\mathrm{g}} \simeq 500\,\mathrm{s}}$ near ${T \simeq 52.3^\circ\text{C}}$ using \eqref{eq:taug}.
These calculations also imply that $R(\tau_{\text{g}}) \simeq 3d$, which means that the crystal mole fraction should be approximately $0.02$ when $t = \tau_{\text{g}}$; comparison with Figure~3d in the main text shows that this estimate of $\tau_{\text{g}}$ is consistent with our diffusion-limited growth data.
Finally, we use this estimate of $\tau_{\text{g}}$ and the temperature-dependent nucleation rates to compute the single-crystal probability via \eqref{eq:isothermalProtocol}.
This prediction compares favorably with the experimental observations shown in Figure~4a of the main text.
We caution that, due to limitations in detecting small crystals, the fraction of single crystals detected in our experiments may be a slight overestimate.
Furthermore, we note that since there is some variation in the colloid concentrations across the droplets, the temperature window within which single crystals can form should be slightly broader than that predicted by our theory.

\subsection{Modeling a time-dependent crystallization protocol}
We now consider a crystallization protocol in which the quench is carried out according to a linear temperature ramp.
Following our approach for calculating $p_{\mathrm{1x}}$ in isothermal experiments, we assume that $\tau_{\text{g}}$ is constant, while $k_{\text{n}}$ is temperature-dependent.
The probability of growing a single crystal using this linear-ramp protocol is
\begin{equation}
  \label{eq:rampProtocol}
  p_{\mathrm{1x}}\left(\left|\frac{dT}{dt}\right|\right) = \left|\frac{dT}{dt}\right|^{-1} \int_{0}^\infty dT\, k_{\mathrm{n}}(T) \exp\left[-\left|\frac{dT}{dt}\right|^{-1} \int_{T - (|dT/dt|)\tau_{\mathrm{g}}}^\infty dT'\, k_{\mathrm{n}}(T')\right],
\end{equation}
where $|dT/dt|$ is the absolute value of the ramp rate.
The predictions of this model compare favorably with experimental measurements using the nominal 0.5\% v/v concentration, as shown in Figure~4b in the main text.
We note that if an annealing protocol uses finite temperature steps instead of a continuous temperature ramp, then the probability of obtaining a single crystal may be lower than that predicted by \eqref{eq:rampProtocol}.
Furthermore, the probability of growing a single crystal using a protocol with finite temperature steps depends on the precise starting temperature in a non-monotonic fashion.
Although our model of a continuous annealing protocol is not intended to account for these effects, we note that \eqref{eq:rampProtocol} remains a good approximation if the temperature steps are small compared to the temperature window over which single crystals can form.

\begin{figure}[h]
  \begin{center}
    \includegraphics{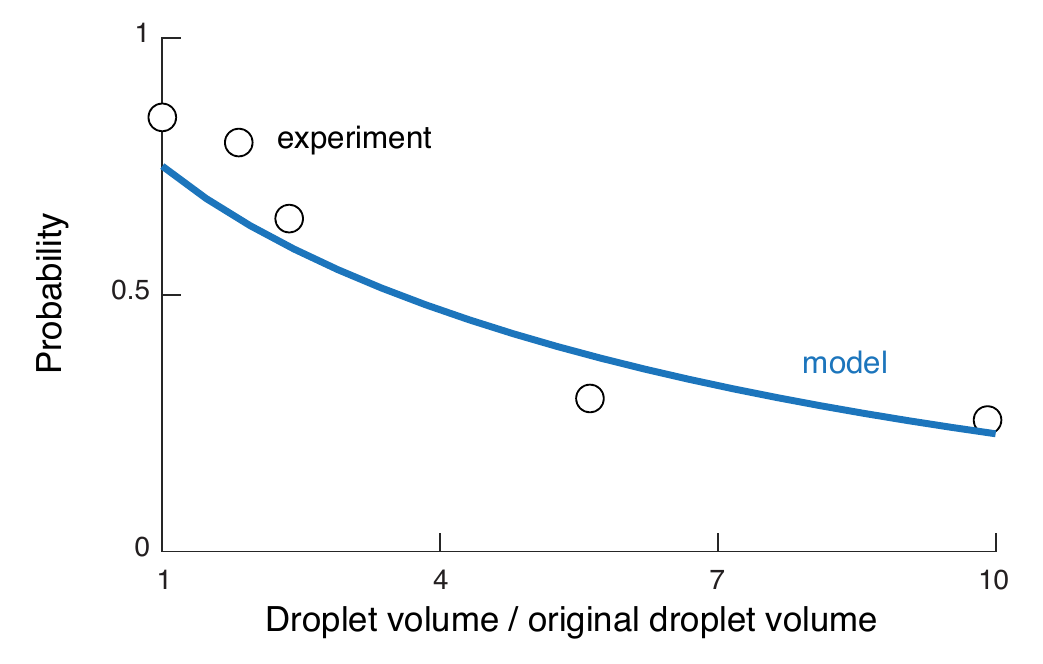}
  \end{center}
  \caption{Comparison of model predictions and experimental results of the probability of forming single crystals for droplets of different volumes. The graphs shows the probability of forming a single crystal within a droplet as a function of the relative droplet volume. The points show experimental measurements and the blue curve shows our model prediction. The original droplet radius is 27 micrometers, which is the same as the droplet radius in Figure 4b--c. Droplets with larger radii have a systematically lower probability of forming single crystals, as captured by our model predictions.  \label{fig:dropletVolumes}}
\end{figure}

As an additional check of our theory, we compare experimental results and model predictions for different droplet volumes.
Intuitively, increasing the droplet volume increases the time that it takes for the nucleus to grow large enough to suppress nucleation elsewhere in the droplet.
To determine how changing the droplet volume affects $\tau_{\text{g}}$, we again assume that suppression of nucleation occurs within the diffusion-limited growth regime of the first nucleus.
This assumption allows us to express $\langle k_{\text{n}} \rangle_t$ in terms of the ratio of $R(t)$ to $R_{\text{drop}}$,
\begin{equation}
  \label{eq:taug-Vdrop}
  \langle k_{\text{n}} \rangle_t = 3 \int_{R(t) / R_{\text{drop}}}^1 dx\,x^2 k_{\text{n}}[\rho(x; R(t)/R_{\text{drop}}), T],
\end{equation}
where $x \equiv r / R_{\text{drop}}$ and, according to \eqref{eq:laplace}, $\rho$ depends only on $x$ and ${R(t) / R_{\text{drop}}}$ if the initial nucleus is assumed to form at the center of the droplet.
Using the fact that the crystal radius grows as $R \sim t^{1/2}$ in the diffusion-limited regime, \eqref{eq:taug} and \eqref{eq:taug-Vdrop} imply the scaling relation ${\tau_{\text{g}} \sim V_{\text{drop}}^{2/3}}$.
With this scaling analysis, we can then use \eqref{eq:rampProtocol} to predict the single-crystal probability as a function of the droplet volume.
The experimental results shown in figure~\ref{fig:dropletVolumes} confirm that, at a fixed ramp rate, increasing the droplet volume indeed lowers the probability of forming a single crystal in a given droplet.
Comparison of our theoretical predictions with these experimental results demonstrates that our theory accurately captures the functional dependence of $\tau_{\text{g}}$, and thus the functional dependence of the single-crystal fraction, on the scaled droplet volume.

\clearpage
\bibliographystyle{Science}
\bibliography{2-SI.bbl}